\documentclass[letterpaper,english,10pt,journal,twocolumn,letterpaper]{IEEEtran}
\usepackage[T1]{fontenc}
\usepackage[latin9]{inputenc}
\usepackage{prettyref}
\usepackage{booktabs}
\usepackage{textcomp}
\usepackage{url}
\usepackage{amsthm}
\usepackage{amsmath}
\usepackage{graphicx}

\makeatletter

\pdfpageheight\paperheight
\pdfpagewidth\paperwidth

\providecommand{\tabularnewline}{\\}

\theoremstyle{plain}
\newtheorem{thm}{\protect\theoremname}
\theoremstyle{definition}
\newtheorem{defn}[thm]{\protect\definitionname}
\theoremstyle{definition}
\newtheorem{example}[thm]{\protect\examplename}
\theoremstyle{plain}
\newtheorem{prop}[thm]{\protect\propositionname}

\usepackage{siunitx}
\usepackage{prettyref}
\usepackage{flushend}

\newcommand{\gtables}[1]{{\noindent\ttfamily\bfseries\begin{minipage}{\linewidth}\centering{}#1\end{minipage}}}

\newrefformat{fig}{Fig.~\ref{#1}}
\newrefformat{tab}{Table~\ref{#1}}
\newrefformat{sec}{Section~\ref{#1}}
\newrefformat{sub}{Section~\ref{#1}}
\newrefformat{ex}{Example~\ref{#1}}

\AtBeginDocument{\setlength\abovedisplayskip{1mm}\setlength\belowdisplayskip{1mm}}

\let\oldcaption = \caption
\renewcommand{\caption}[1]{\oldcaption{#1}\vspace{-3mm}}

\@ifundefined{showcaptionsetup}{}{%
 \PassOptionsToPackage{caption=false}{subfig}}
\usepackage{subfig}
\makeatother

\usepackage{babel}
\providecommand{\definitionname}{Definition}
\providecommand{\examplename}{Example}
\providecommand{\propositionname}{Proposition}
\providecommand{\theoremname}{Theorem}

\begin{document}

\title{Air Dominance in Sensor Networks:\\
Guarding Sensor Motes using Selective Interference}

\author{Matthias Wilhelm, Ivan Martinovic, Jens B.~Schmitt, and Vincent
Lenders\thanks{%
Matthias Wilhelm and Jens B.~Schmitt are with the Department of Computer Science, University of Kaiserslautern, Paul-Ehrlich-Str.~36, 67663 Kaiserslautern, Germany. E-mail: \{wilhelm, jschmitt\}@cs.uni-kl.de%
}%
\thanks{Ivan Martinovic is with the Department of Computer Science, University of Oxford, Parks Road, Oxford OX1 3QD, UK. E-mail: ivan.martinovic@cs.ox.ac.uk}%
\thanks{Vincent Lenders is with armasuisse Science and Technology, armasuisse, 3602 Thun, Switzerland. E-mail: vincent.lenders@armasuisse.ch}}
\maketitle
\begin{abstract}
Securing wireless sensor networks (WSNs) is a hard problem. In particular,
network access control is notoriously difficult to achieve due to
the inherent broadcast characteristics of wireless communications:
an attacker can easily target any node in its transmission range and
affect large parts of a sensor network simultaneously. In this paper,
we therefore propose a distributed guardian system to protect a WSN
based on physically regulating channel access by means of selective
interference. The guardians are deployed alongside a sensor network,
inspecting all local traffic, classifying packets based on their content,
and destroying any malicious packet while still on the air. In that
sense, the system tries to gain ``air dominance'' over attackers.
A key challenge in implementing the guardian system is the resulting
real-time requirement in order to classify and destroy packets during
transmission. We present a USRP2 software radio based guardian implementation
for IEEE 802.15.4 that meets this challenge; using an FPGA-based design
we can even check for the content of the very last payload byte of
a packet and still prevent its reception by a potential victim mote.
Our evaluation shows that the guardians effectively block 99.9\,\%
of unauthorized traffic in 802.15.4 networks in our experiments, without
disturbing the legitimate operations of the WSN.\end{abstract}

\section{Motivation}

Wireless Sensor Networks (WSNs) are extending their application scope
from industrial monitoring and location tracking to more personal
and assistant technologies, such as in health care \cite{MediSens},
assisted living \cite{AssistedL}, and home energy saving applications
\cite{EnergyAudit,Thermostat}.  ZigBee-enabled devices such as door
locks, occupancy sensors, panic buttons and electrical sockets are
already available as low-cost consumer electronics ready to be deployed
in users' residences. Imagine an emergency scenario where a gas leakage
detector rises an alarm or a panic button is pressed, and since an
occupancy sensor reports an occupied room, the door lock system decides
to unlock and provide emergency exits. While such a scenario is a
perfect motivation for using WSN technologies, it also provides an
attractive playground for an attacker.

In contrast to wired networks where physical control of traffic is
inherently given, wireless networks are open by nature. For this reason,
both IEEE 802.15.4 (at the link-layer) and ZigBee (at the upper layers)
define conventional security services for frame protection, device
authorization, key distribution, and key establishment. However, they
also take into account restrictions of battery-powered, performance-limited
and low-cost devices and offer tradeoffs between resource requirements
and security objectives, depending on the particular application scenario.
For example, according to the IEEE 802.15.4 standard, there are three
security modes: \emph{(i)} no security, \emph{(ii)} access control
lists (ACLs) based on a source address, and \emph{(iii)} secured mode,
offering a choice of strong security suites such as 128\,bit AES-CCM.
From a security perspective, only the latter option offers protection
in an adversarial setting. Similarly, the ZigBee-2007 specification
describes key management and key exchange methods. It specifies three
types of keys: \emph{(i)} master key, used as an initial shared secret
to generate link keys, \emph{(ii)} link key, dynamically generated
secret keys shared only between two devices, and \emph{(iii)} network
key, a global secret key shared among all WSN devices. Yet, the master
and link keys are optional. Hence, it is realistic to assume that
the security of standard ZigBee networks may reside in only a single
shared key with the obvious risk that the capture of a single device
and extraction of the secret key could jeopardize the security of
the whole network.  Along these lines, a recently available security
analysis toolbox called KillerBee \cite{Killerbee} offers a set of
attack vectors, such as Over-the-Air (OTA) key sniffing, MAC address
manipulation, key extraction from memory, and denial-of-service attacks
based on flooding WSNs with memory-consuming association requests.

While it is understandable that securing WSNs includes optimizing
various tradeoff ``knobs,'' we are concerned that this complex task
together with the distributed nature of WSNs will result in inconsistent
security configurations, misconfigured clients, complex key revocations,
and often in a complete resignation from implementing any serious
security measure. To overcome these problems, we describe our design
objectives as a ``wish list'' of features that we believe should
be part of an effective and practical solution:
\begin{itemize}
\item \textbf{Central control }for WSN security policy enforcement
\item \textbf{Outsourcing security costs} from sensor devices
\item \textbf{Transparency} with respect to existing WSN protocols
\item \textbf{Generic} \textbf{protection} supporting different communication
protocols
\item \textbf{On-demand} \textbf{security}, i.e., paying security overheads
only during attacks.
\end{itemize}
In wired networks, many of these properties are found within the concept
of \emph{network firewalls}. Firewalls are store-and-forward devices
located at the edges of networks, controlling the access to the networks
they are to protect, effectively defining a trust boundary between
the inside and outside world by blocking any untrusted traffic from
reaching the protected devices. Network administrators are generally
very much in favor of firewalls as they enable a \textit{\emph{central
control}}\textit{ }of policy enforcement. This is, in principle, achieved
in a\textit{ }\textit{\emph{transparent}}\textit{ }\textit{\emph{way}},
without necessitating changes to existing protocols or host configurations.
A firewall's actions are usually specified by \textit{\emph{generic}}
policies defined by filtering rules. Hence, it is hard to deny that
the concept of enforcing security policies by blocking unwanted traffic
before it reaches the clients could also be attractive in supporting
a practical approach to WSN security. The problem is, however, that
in contrast to wireline networks, the broadcast nature of the wireless
channel does not provide any physical separation of the traffic and
thus setting up a boundary between inside and outside world is much
harder to achieve, especially when considering mobile nodes. Consequently,
preventing a packet from being received cannot be based on simple
and silent dropping at a store-and-forward device but requires a different
mechanism.

Recently, the idea of creating intentional radio frequency interference
and turning it against an adversary has shown to be a valuable alternative
(or addition) to conventional security protocols. Such a ``friendly
jamming'' paradigm has been proposed to, e.g., protect implanted
medical devices \cite{Heartbeats,IMDGuard} or block suspicious transmissions
in WSNs \cite{JamGood} (more details on related work are given in
\prettyref{sec:RelatedWork}). Yet, these protocols use radio interference
either as a proactive countermeasure, i.e., jamming all communication
on a certain frequency band, or demand additional higher-layer communication
protocols. Hence, the existing approaches violate our design objectives
of transparency, generic protection, and on-demand security.

In this work, we describe a guardian system operating at the physical
layer that is able to inspect packets (including both packet header
and payload) that are already on the air. Then, using a predefined
security policy, it classifies a packet in real-time and, in case
of policy violations, generates a ``surgically'' precise interference
that introduces bit errors at legitimate receivers such that the malicious
packet is discarded. The security policies are centrally defined and
enforced ($\rightarrow$~central control) and wireless sensors are
not required to implement complex security mechanisms such as per-device
packet filtering \cite{AEGIS} ($\rightarrow$~outsourcing security
costs). Since blocked packets are essentially not received by the
sensor devices, the WSN is not aware of this countermeasure, i.e.,
what is correctly received in the network has already been verified
($\rightarrow$~transparency). And because this protection method
is implemented on the physical layer, it is independent of any higher-layer
protocols ($\rightarrow$~generic protection). Finally, the reactive
behavior of the guardian system only activates if packets that violate
security policies are on the air, which means that there is presently
an actual attack against the WSN ($\rightarrow$~on-demand security).
The system is implemented on the widely used software radio platform
USRP2, which also facilitates its usage as an open experimental platform.
Besides meeting the design goals, we make the following contributions:
\begin{itemize}
\item We present an FPGA-based system implementation with an overall reaction
delay of \SI{39}{\micro\second}, allowing a per-packet classification
based on packet contents, up to the very last payload byte. The system
is very configurable, supporting the definition of access policies
in the style of \texttt{iptables}.
\item Our overall system evaluation shows that even with very limited interference
duration (\SI{26}{\micro\second} per packet) and large distances
(up to 18\,m), the system is able to destroy 99.9\,\% of malicious
packets in our experiments.
\item The effectiveness of our protection approach is validated in an IEEE
802.15.4 sensor network using the open-source toolbox KillerBee (as
a realistic attacker).
\end{itemize}
The rest of the paper is structured as follows: In \prettyref{sec:Concept},
we describe the key concepts of our guardian system in more detail.
A theoretical protection analysis of these concepts directly follows
in \prettyref{sec:Analysis}. \prettyref{sec:Impl} presents the most
interesting details of the guardian node implementation. A comprehensive
testbed evaluation of critical system aspects is provided in \prettyref{sec:Evaluation}.
In \prettyref{sec:Apps}, we illustrate the potential value of the
guardian system in several real-world application scenarios. \prettyref{sec:Discussion}
provides a discussion of limitations, some open issues, and future
research opportunities. In \prettyref{sec:RelatedWork}, we thoroughly
review related work and conclude the paper in \prettyref{sec:Conclusion}.

\section{Air Dominance Concept}

\label{sec:Concept}
\begin{figure*}
\centering \includegraphics[width=0.85\textwidth]{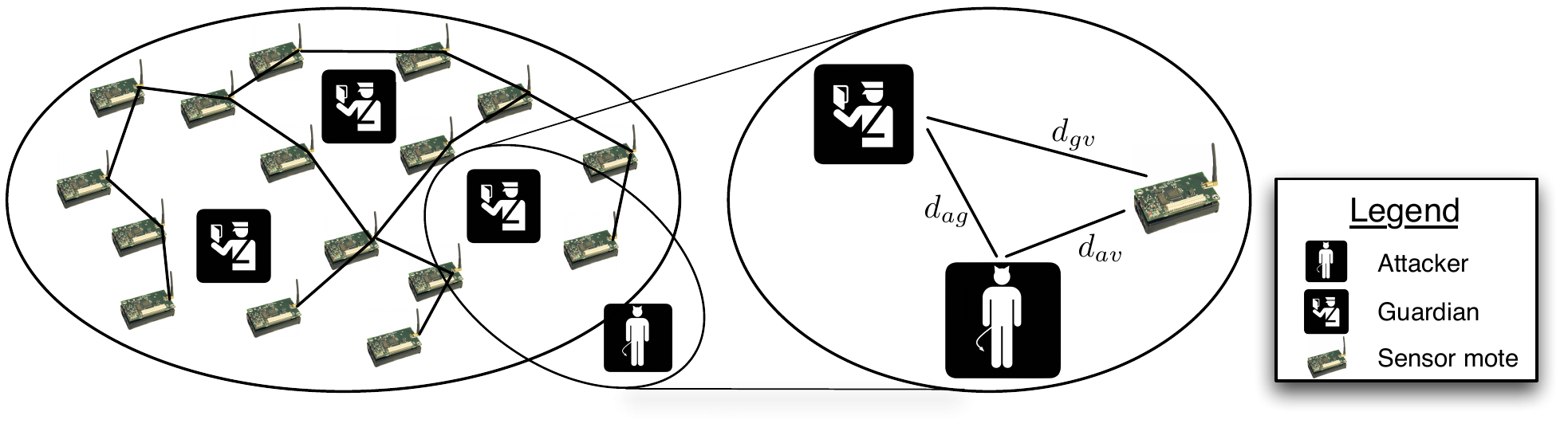}\caption{\label{fig:SensorNetwork}Sensor network deployment with co-located
guardian nodes. In our protection analysis (Section~\ref{sec:Analysis}),
we use the distances on the right-hand side to show that the maximal
attacker distance $d_{av}$ is drastically reduced by the guardian
system.}
\end{figure*}
In this section, we outline how the physical layer protection of our
guardian system operates in general, and what protection properties
are achieved by it.

\subsection{System Model and Assumptions}

We consider a wireless sensor network scenario with three types of
devices (see \prettyref{fig:SensorNetwork}):
\begin{enumerate}
\item Wireless sensor nodes that perform a distributed sensing application;
these nodes are the target of attacks and hence are referred to as
\emph{victim} nodes~$v$.
\item \emph{Attacker} nodes~$a$ that attempt to send malicious packets
to victim nodes~$v$ to disrupt their intended network operation.
\item \emph{Guardian} nodes~$g$ that want to prevent victim nodes~$v$
from receiving malicious packets from attackers~$a$.
\end{enumerate}
The sensor nodes communicate using IEEE 802.15.4 on the 2.4\,GHz
band and we assume that all nodes send and receive on the same channel.
We further assume that the victim sensor nodes~$v$ act in compliance
to the IEEE 802.15.4 standard and that they discard received packets
for which the CRC checksum is incorrect. The CRC checksum is a 16\,bit
field calculated over a packet's payload and headers. The CRC checksum
is erroneous when a packet has at least one symbol error (we do not
consider error coding mechanisms such as \cite{BuzzBuzz}). We further
assume that the victim nodes communicate without header encryption,
the guardian nodes can therefore eavesdrop and decode any transmitted
packets. This is no limitation of our concept for two reasons: First,
header encryption is generally not used in wireless networks because
all messages must be received, decrypted, and checked for integrity
using all available link keys, which is extremely inefficient; the
IEEE 802.15.4 standard only considers \emph{data} confidentiality
as a security service \cite[$§$5.5.6]{IEEE802.15.4-2006}. Second, even
if header encryption is used guardians can be given access to the
network's cryptographic material because they are part of the network
infrastructure. Then a guardian can decrypt packets during transmission
(there are efficient FPGA implementations of AES available \cite{AESFPGA})
and block them, given the packets are longer than one AES block.

\subsection{Guardian Operation}

Guardian nodes are responsible for blocking unauthorized traffic sent
by the attacker nodes to the victim nodes. Each guardian continuously
performs the following steps:
\begin{enumerate}
\item Monitor the RF channel for suspicious transmissions.
\item Decide whether a detected packet is allowed or not (before it is fully
received by sensor motes).
\item If the packet is not allowed, briefly emit interference to destroy
the packet.
\end{enumerate}
Packets are destroyed in Step~3) by exploiting the CRC mechanism
of IEEE 802.15.4. Since the CRC of packets at victim nodes is erroneous
when at least one symbol is wrong, packets that are hit strong enough
by interference from a guardian node are discarded by the victim receivers.

To react quickly, the guardians operate autonomously. This means that
a packet may be hit by multiple guardians when they detect that a
packet that is not allowed. We see in our evaluation ($\rightarrow$~\prettyref{sub:JamPerf})
that this independent operation of the guardian nodes is beneficial,
particularly in environments where individual guardians may not reliably
overhear all packets. The drawback of this approach is, of course,
that we introduce more interference than necessary, which could be
problematic for the operation of the sensor network or other wireless
networks in the same frequency band. One system design goal for the
guardians is therefore to make this interference as ``friendly''
as possible by using very short periods of interference and using
waveforms that are effective at the target sensor nodes but negligibly
disturbing for other wireless networking technologies operating in
the vicinity. This is important because an unintentional derogation
of co-existing networks could prevent a certification of guardian
devices or reduce the acceptance of their deployments; we discuss
legal aspects in \prettyref{sub:Legal}.

\subsection{Guardian Design Considerations}

To be effective the guardians should reliably detect all packets on
the air and interfere with all unauthorized ones. At the same time,
the interference should be limited in order to minimize possible effects
on concurrent communications on the same frequency band. We next analyze
the design tradeoffs for the implementation of the guardians' receive
and blocking primitives.

\subsubsection{Reflections on Guardian Detection Sensitivity}

\label{sub:Sensitivity}In order to correctly detect and demodulate
a packet at a receiver, the incoming signal power at the antenna must
be greater than the receiver sensitivity. This sensitivity represents
the minimum signal power at the antenna that results in a specified
packet error performance. The receiver sensitivity is given by $S=N_{T}N_{F}\textrm{SN\textrm{R}}_{\textrm{min}}$
\cite{Adamy01}, where $N_{T}$ is the thermal noise, $N_{F}$ is
the noise figure of the particular receiver and $\textrm{SNR}_{\textrm{min}}$
a modulation-specific threshold for the minimum required signal-to-noise
ratio. For IEEE 802.15.4-compliant receivers with O\nobreakdash-QPSK
modulation on the 2.4\,GHz band, optimal coherent detection, and
a maximum packet error rate of 1\,\%, the theoretical sensitivity
limit is $S_{\textrm{min}}=-112.2$\,dBm. However the standard only
demands a sensitivity of at least $-85$\,dBm, and commercial radio
modules exhibit sensitivities ranging between $-92$\,dBm to $-110$\,dBm.%
\footnote{An online comparison of various commercial receiver implementations
with links to the corresponding data sheets is available at \url{http://en.wikipedia.org/wiki/Comparison_of_802.15.4_radio_modules}%
}

Our analysis in \prettyref{sec:Analysis} shows that the guardians
benefit from a high sensitivity. Next, we discuss several approaches
to help increasing the guardian's detection performance. Lanzisera
and Pister \cite{RXLimits} show that 6.6\,dB are generally sacrificed
by design decisions, but 5.8\,dB of this can easily be reclaimed
with improved receiver implementations. A better sensitivity can also
be provided by using low noise components and more efficient demodulating
schemes.

The sensitivity of guardian nodes can even be designed to go beyond
the theoretical sensitivity limit of $-112.2$\,dBm of regular receivers
in sensor nodes. At $-112.2$\,dBm, an ideal IEEE 802.15.4 receiver
will exhibit less than 1\,\% packet errors, however, guardian nodes
may exploit the fact that it is not necessary to receive all bits
in a packet correctly to successfully classify it. Bit errors that
occur outside the portion of the packet that matches a rule have no
effect on the detection and may hence be tolerated by guardian nodes.
Even bit errors that occur within the portion of the packet of a rule
may be tolerated at the cost of occasionally destroying a packet that
does not match the specified rule. By tolerating up to 10\,\% BER,
the sensitivity can be improved by more than 11\,dB \cite{RXLimits}.
Significant detection improvements are hence realizable. The sensitivity
gain of tolerating bit errors is substantial while causing only few
false positives. For example, considering a blocking rule that is
matched to a 32\,bit address field, if we tolerate a 2\,bit error
(6.25\,\% BER) and assuming a uniform address distribution, the false
positive rate remains below 0.00002\,\% while the sensitivity is
increased by 10\,dB. However, this analysis may be affected by other
rules in the active rule set if they consider the same portions of
the packet. We assume in this work that modest improvements are possible
and leave a detailed analysis of achievable sensitivity gains to future
work.

\subsubsection{Reflections on Guardian Interference}

To achieve ``friendly'' interference, the duration of the blocking
signals should be as short as possible. In IEEE 802.15.4, one erroneous
symbol already leads to a wrong CRC checksum, and the packet at the
victim node is discarded. Hence, a guardian who detects an unauthorized
packet ideally needs to interfere with that packet for the duration
of a single symbol (\SI{16}{\micro\second}). We investigate in \prettyref{sub:JamDuration}
that, under real-world settings, it requires a slightly higher duration
of selective guardian interference to effectively block all undesired
packets. However, this short interference period is still limited
and negligible compared to the signal duration of the packet emitted
by the attacker. A wireless sensor network will hence not be significantly
impacted since IEEE 802.15.4 nodes perform carrier sensing and should
not transmit concurrently with the attacker anyway.

Another aspect of friendly interference relates to the choice of the
interference waveform. It is well known that certain waveforms are
more effective at causing bit errors than others \cite{PoiselJamming}.
The effectiveness of the interference waveform depends on the particular
modulation scheme and the filters at the receivers. Advantageously,
wireless sensor nodes are generally not designed to suppress in-band
interference well. They only low-pass filter the baseband signal and
send the output directly to the demodulator \cite{JamFilters}. We
see in our experimental evaluation in \prettyref{sub:JamDuration}
that a continuous sine wave, a waveform that is generally known as
being ineffective, happens to destroy packets even more effectively
in IEEE 802.15.4 with the same power budget than band-limited white
noise or modulated signals. Using a very narrowband waveform like
this provides the advantage that other technologies like WLAN are
able to efficiently suppress this kind of interference \cite{Karhima04}
and hence remain unaffected by the guardians.

\section{Protection Analysis}

\label{sec:Analysis}
\begin{table}
\centering %
\begin{tabular}{cl}
Term & Description\tabularnewline
\hline 
$d_{av}$ & Distance between the attacker and the victim\tabularnewline
$d_{ag}$ & Distance between the attacker and the guardian\tabularnewline
$d_{gv}$ & Distance between victim and the guardian\tabularnewline
$P_{a}$ & Radiated power by the attacker\tabularnewline
$P_{g}$ & Radiated (interference) power by the guardian\tabularnewline
$S_{v},S_{g}$ & Receiver sensitivity of victim and guardian\tabularnewline
$\gamma_{\textrm{SIR}}$ & Signal-to-interference threshold for reception\tabularnewline
$\alpha$ & Path loss coefficient\tabularnewline
$L\left(d\right)$ & Path loss and fading function of distance $d$\tabularnewline
$\mathcal{A}_{\textrm{force}},\mathcal{A}_{\textrm{stealth}}$  & Attackers using a brute-force or stealthy strategy\tabularnewline
$\mathcal{R}_{\mathcal{A}}$ & Resulting attack range, performance metric for attackers\tabularnewline
\end{tabular}

\caption{\label{tab:Notation}Notation used in the protection analysis.}
\end{table}
In this section, we analyze the nature of protection offered by our
system of distributed guardians. For ease of exposition, we deliberately
use simple propagation and receiver models to deliver the key insights
on the concept (we use log-distance path loss and an SINR-based receiver
model). The notation used in this section is summarized in \prettyref{tab:Notation}
for convenience.

\subsection{Protection Constraints}

For a guardian~$g$ to protect a victim node~$v$ from an attacker~$a$,
the guardian must be able to detect the signals from the attacker
and interfere with the attacker's signal at the victim node such that
the packet is discarded. This leads to the following two necessary
conditions for guardian protection.

\textit{Condition 1 (Detection):} Let $P_{a}$ be the emitted signal
power of the attacker and $L\left(d_{ag}\right)$ be the path loss
and fading between the attacker and the guardian. The guardian is
able to detect the malicious signals iff
\[
P_{a}L\left(d_{ag}\right)\geq S_{g},
\]
where $S_{g}$ is the sensitivity of the guardian.

\textit{Condition 2 (Destruction): }Let $P_{a}$ be the emitted signal
power of the attacker, $P_{g}$ be the emitted power of the guardian's
interference signal, and $L\left(d_{av}\right)$, $L\left(d_{gv}\right)$
be the path loss between attacker--victim and guardian--victim, respectively.
The guardian is able to destroy the packet at the victim if the signal-to-interference
ratio is below a certain threshold
\[
\frac{P_{a}L\left(d_{av}\right)}{P_{g}L\left(d_{gv}\right)}<\gamma_{\textrm{SIR}},
\]
where the threshold $\gamma_{\textrm{SIR}}$ is determined by the
modulation scheme, the interference waveform and how well the victim
node is able to suppress such interference. If the interference waveform
is zero-mean white gaussian noise, the threshold $\gamma_{\textrm{SIR}}$
is between 0\,dB and 3\,dB according to the IEEE 802.15.4 standard
\cite{IEEE802.15.4-2006}. More effective waveforms by the guardians
leads to higher values of $\gamma_{\textrm{SIR}}$. We evaluated this
value for the MICAz platform with several waveforms in \prettyref{sub:JamDuration}
and find an additional gain of 3--5\,dB in jamming effectiveness.

A sensor network remains protected if there exists at least one guardian
that fulfills these two conditions for each attacker and victim node
pair. In the remaining of this section, we discuss that these conditions
are generally easy to fulfill as there exists a large asymmetry between
the capability requirements for attackers and guardians.

\subsection{Attacker Models}

The goal of the attacker is to inject packets of its choice into the
protected network. We consider two attacker strategies against the
guardians:

\textit{Brute-force attacker $\mathcal{A}_{\textrm{force}}$}\textit{\emph{:}}
the attacker tries to overcome the guardian's interference by using
a large transmission power $P_{a}$:
\[
\frac{P_{a}L(d_{av})}{P_{g}L(d_{gv})}>\gamma_{\textrm{SIR}}.
\]

\textit{Stealthy attacker $\mathcal{A}_{\textrm{stealth}}$}\textit{\emph{:}}
the attacker tries to choose a small transmission power $P_{a}$ such
that the injected packet is received only by the victim but is not
detected by the guardian. The attack is successful iff
\[
\begin{array}{ccc}
P_{a}L(d_{av})\geq S_{v} & \mathrm{and} & P_{a}L(d_{ag})<S_{g}.\end{array}
\]

While the first strategy is easy to implement, it also leads to a
steeply increased energy cost of the attacker (an analysis for 802.15.4
is given in Proposition~\ref{prop:EnergyCost}). The second strategy
is more challenging to implement in practice as very fine-grained
power control and a clever positioning of the attacker is required
to perform the attack successfully.

\subsection{Attacker--Guardian Asymmetry}

To illustrate the asymmetry between an attacker attempting to inject
malicious packets and the guardian protecting the victim nodes, we
consider a log-normal path loss model $L(d)=L_{0}d^{-\alpha}$, with
reference path loss~$L_{0}$, distance~$d$, and path loss coefficient~$\alpha$.
We use the following metric to capture the asymmetry between an attacker
and a guardian:
\begin{defn}
[Attack Range]The (worst-case%
\footnote{This notion of attack range is under worst-case assumptions where
attacker, victim, and guardian are on one line with the victim in
the middle, maximizing the distance between attacker and guardian
$d_{ag}$ for a given attacker--victim distance $d_{av}$.%
}) attack range $\mathcal{R}$ is the maximum distance $d_{av}$ between
attacker~$a$ and victim node~$v$ such that an attack still succeeds.
\end{defn}
An attacker wants the attack range to be as large as possible such
that it may launch the attack from arbitrary locations and still remain
undetected in the physical world. A large attack range further makes
an attacker powerful as it can attack more nodes from a single location.
From the perspective of the network, a small attack range is desired
as it forces an attacker to expose itself in the physical world and
minimizes the number of victim nodes that an attacker may attack simultaneously.
If the attacker is constrained in terms of reachable locations, e.g.,
when the attacker can only attack from outside a building, the number
of sensor motes in its attack range may also be zero for all reachable
attack positions, effectively thwarting the attack.
\begin{example}
\label{ex:FullRange}Without a guardian, the attack range $\mathcal{R}_{\mathcal{A}_{\textrm{force}}}$
of an attacker $\mathcal{A}_{\textrm{force}}$ using a powerful COTS
transmitter ($P_{a}=20$\,dBm) and MICAz victims ($S_{v}=-94$\,dBm)
under the log-normal model with path loss parameters specified in
the IEEE 802.15.4 standard \cite[Annex E.5.3]{IEEE802.15.4-2006}%
\footnote{We converted the parameters given in the standard to our non-logarithmic
model: reference distance $d_{0}=8$\,m, $\alpha=3.3$, path loss
at $d_{0}$ is $L_{0}=d_{0}^{\alpha}10^{58.5/10}$. Later derivations
use distance ratios and are thus independent of the choice of $L_{0}$.%
} is $\mathcal{R}=384.51$\,m.
\end{example}
The effect of an active guardian is to considerably reduce the attack
range. We derive bounds on the attack range for both attacker models
$\mathcal{A}_{\textrm{stealth}}$ and $\mathcal{A}_{\textrm{force}}$
in the following.
\begin{prop}
The attack range $\mathcal{R}_{\mathcal{A_{\textrm{stealth}}}}$ of
attacker $\mathcal{A}_{\textrm{stealth}}$ is bounded by $d_{vg}/\left(\sqrt[\alpha]{S{}_{v}/S_{g}}-1\right)$
if a guardian is present.\end{prop}
\begin{IEEEproof}
To prevent detection, the attacker must satisfy two conditions simultaneously:
its received signal power at the victim must be above the sensitivity
level of the victim and below the sensitivity of the guardian:
\begin{gather}
P_{a}d_{av}^{-\alpha}\geq S_{v}\label{eq:VHear}\\
P_{a}d_{ag}^{-\alpha}<S_{g}\label{eq:WNotHear}
\end{gather}
Following (\ref{eq:VHear}) and (\ref{eq:WNotHear}), the attacker
must choose a power $P_{a}$ that satisfies $d_{av}^{\alpha}S_{v}\leq P_{a}<d_{ag}^{\alpha}S_{g}$.
Thus, such a $P_{a}$ only exists iff $d_{av}^{\alpha}S_{v}<d_{ag}^{\alpha}S_{g}$.
Using this condition and the relation $d_{ag}\leq d_{av}+d_{gv}$
(by triangle inequality), we find that the attacker $\mathcal{A}_{\textrm{stealth}}$
can only remain undetected if its distance to $v$ satisfies 
\begin{equation}
d_{av}<\frac{d_{vg}}{\sqrt[\alpha]{S_{v}/S{}_{g}}-1}.
\end{equation}
The bound is independent of the attacker's transmit power.
\end{IEEEproof}
The asymmetry with the attacker model $\mathcal{A}_{\textrm{stealth}}$
is provided by the sensitivity ratio $S_{v}/S_{g}$ (as discussed
in \prettyref{sub:Sensitivity}): the higher this ratio, the smaller
is the attack range $\mathcal{R}_{\mathcal{A}_{\textrm{stealth}}}$.
Additionally, this range only applies if the attacker maximizes its
distance to the guardian, such that it has to find a position where
the attack is feasible first. The following example applies typical
values to illustrate how the attack range is considerably reduced
compared to \prettyref{ex:FullRange} where no guardian is present.
\begin{example}
In \prettyref{sub:App--KillerBee}, we present an indoor application
scenario with MICAz sensor motes and a KillerBee attacker. Assuming
a near-optimal receiver at the guardian with a sensitivity of $-110$\,dBm
($-112.2$\,dBm is the theoretical sensitivity limit of an ideal
802.15.4 receiver), tolerating a limited number of bit errors and
hence a non-zero false positive rate with a conservative improvement
of 6\,dB to keep the false positive rates low (say less than 0.001\,\%),
the sensitivity of the guardian becomes $S_{g}=-116$\,dBm, resulting
in a ratio $S_{v}/S_{g}$ of 22\,dB. With a maximal protection distance
$d_{gv}=10$\,m and $\alpha=3.3$, the attack range decreases to
$\mathcal{R}_{\mathcal{A}_{\textrm{stealth}}}=2.75$\,m, which is
a reduction by a factor of 140 compared to the case where no guardian
is present.\end{example}
\begin{prop}
The attack range $R_{\mathcal{A_{\textrm{force}}}}$ for attacker
model $\mathcal{A}_{\textrm{force}}$ is bounded by $\sqrt[\alpha]{P_{a}/(\gamma_{\textrm{SIR}}P_{g})}d_{gv}$.\end{prop}
\begin{IEEEproof}
If the attack signal is detected, a brute-force attacker has still
the opportunity to overcome the interference signal if 
\[
\frac{P_{a}d_{av}^{-\alpha}}{P_{g}d_{gv}^{-\alpha}}\geq\gamma_{\textrm{SIR}}.
\]
Consequently, the attacker must choose its distance such that 
\begin{equation}
d_{av}\leq\sqrt[\alpha]{\frac{P_{a}}{\gamma_{\textrm{SIR}}P_{g}}}d_{gv}.
\end{equation}
The attacker is assumed to be power-limited, such that this bound
exists.\end{IEEEproof}
\begin{example}
In \prettyref{sub:App--Revocation}, a sensor network with several
MICAz motes captured by an attacker is considered, which our guardian
system effectively disconnects from the network. MICAz motes (like
many other COTS transmitters) have a maximum output power of $P_{a}=0$\,dBm,
the transmit power of a guardian node is $P_{g}=20$\,dBm, and the
interference waveform being used is assumed as $\gamma_{\textrm{SIR}}\approx3$\,dB.
With a maximal protection distance $d_{gv}=10$\,m, the attack range
$\mathcal{R}_{\mathcal{A}_{\textrm{force}}}$ of the brute-force attacker
is brought down to 0.70\,m.
\end{example}
Last but not least, a brute-force attacker also has to pay a price
in terms of energy investment to successfully mount its attack.
\begin{defn}
[Energy cost]The energy cost of an attacker $\mathcal{A}_{\textrm{force}}$
is the ratio of the energy required by the attacker to send a packet
over the energy required by the guardian to block the packet. \end{defn}
\begin{prop}
\label{prop:EnergyCost}The energy cost is 
\begin{equation}
\frac{P_{a}d_{av}^{-\alpha}\ell\cdot\SI{32}{\micro\second}}{P_{g}d_{gv}^{-\alpha}t_{\textrm{interfere}}},
\end{equation}
where $\ell$ denotes the packet length in bytes and $t_{\textrm{interfere}}$
the duration of the interference signal from the guardian.\end{prop}
\begin{example}
If, for ease of exposition, we assume that the attacker and the guardian
emit at the same power and are at the same distance from the victim,
the (received) power ratio $P_{a}d_{av}^{-\alpha}/P_{g}d_{gv}^{-\alpha}$
is equal to one. Considering a typical IEEE 802.15.4 packet size of
32\,bytes and a necessary interference duration of $t_{\textrm{interfere}}=\SI{16}{\micro\second}$
to destroy one symbol, the energy cost is 64. In other words, the
guardian needs to invest 64 times less transmit energy to destroy
a packet than the attacker needs to invest to transmit the packet
to the victim.
\end{example}
This observation shows again an asymmetry in the required energy between
an attacker and a guardian. While an attacker might in some cases
still be able to slip through the protection of the guardians, the
effort required to do so is considerably higher than the effort required
by the guardians. This asymmetry is precisely what we refer to as
\emph{air dominance}, where guardians are able to effectively and
efficiently control and block undesired transmissions from attackers
before they can reach the network. 

\section{Guardian Implementation}

\label{sec:Impl}
\begin{figure*}
\centering \includegraphics[width=0.8\textwidth,height=1in]{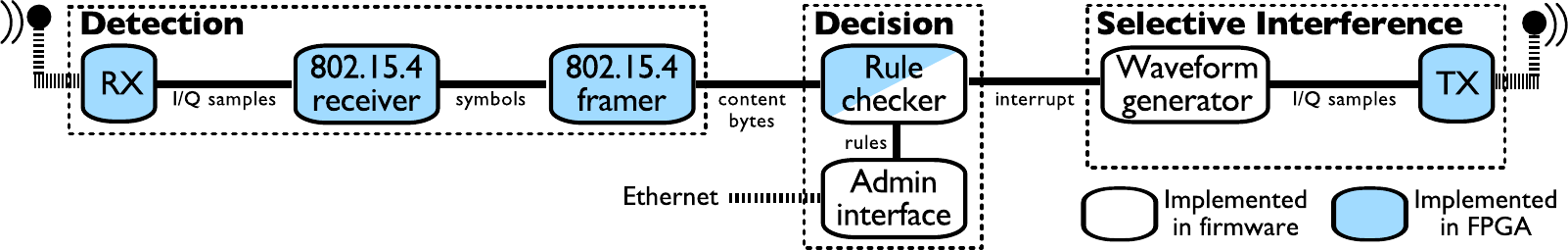}

\caption{\label{fig:Workflow}Component diagram of the guardian implementation.
We consider two different rule checker implementations (firmware\slash{}FPGA).}
\end{figure*}
After introducing the concept, we describe the implementation of our
system that performs the detection, classification, and destruction
of malicious packets. The system design is driven by the need for
low-latency operation, as the reaction time of the system must be
smaller than the duration of a packet to both classify and destroy
it. And because the classification is based on packet content, the
allowed reaction time should be only a fraction of a packet's duration,
i.e., in the order of tens of microseconds.

We implement the guardian nodes on the USRP2 software-defined radio
(SDR) platform. The SDR paradigm allows full physical layer access
by the guardian; in contrast, off-the-shelf receivers do not allow
low-latency access to detected symbols, nor do they offer freedom
in choosing the transmitted interference waveform. However, these
factors are crucial to enable a deep look into packets and to use
the most effective interference waveform with a given power budget.
For example, related reactive jamming systems implemented on sensor
motes \cite{JamGood,ReactiveTestbed,TwoRadio} do not allow to classify
based on content and to destroy packets during transmission. Still,
an implementation on the USRP2 is not straightforward because the
time to classify and intercept a packet is below \SI{500}{\micro\second},
which forces tight timing constraints on the system. In the common
host-based architecture for software radio using GNU Radio, where
digital samples are forwarded to a host via Ethernet and signal processing
is performed on the host CPU, large and non-deterministic latencies
prevent to meet real-time requirements \cite{SDRMAC}. In our implementation,
we fulfill the requirements of real-time detection and subsequent
destruction of a packet on the air; we design the guardian as a stand-alone
system, completely implemented on the USRP2's FPGA and on-board micro-controller
(our previous work is based on the same framework \cite{ReactThreat,WMSL11-3}).
The system is optimized for speed to reach reaction times of tens
of microseconds to allow access to the full payload and still reliably
destroy the packet. The system components are depicted in \prettyref{fig:Workflow},
the workflow of the system is (from left to right): \emph{(i)} detecting
and demodulating relevant transmissions, \emph{(ii)} forming decisions
based on the packet content, and \emph{(iii)} interfering with malicious
packets. The system currently supports the 2.4\,GHz PHY of IEEE 802.15.4.
Next, we provide a brief description of each functional component.

\subsection{Detection Subsystem}

This subsystem continuously scans the RF medium to detect any packet
that might be received by the network and receives and delivers the
packet's content to the subsequent decision subsystem. It consists
of a receiver optimized for speed that synchronizes to packets and
demodulates the contained bytes, and a framer that interprets the
bytes according to the IEEE 802.15.4 standard, providing access to
header fields and payload bytes.

\textbf{802.15.4 receiver.} We implemented an IEEE 802.15.4 receiver
using coherent O\nobreakdash-QPSK demodulation and a correlating
direct sequence de-spreader to recover symbols in FPGA logic. The
de-spreader removes the direct sequence spreading code used in the
2.4\,GHz PHY. The receiver directly operates on the stream of complex
(I/Q) samples coming from the USRP2's analog-to-digital converter:
it first synchronizes with the incoming preamble, then detects each
detected symbol (4 bit of information), and finally delivers it to
the framer. The FPGA implementation ensures that the detection latency
is limited, we measured a delay in the receiver below \SI{4}{\micro\second}
from the presence of the signal on the channel to the time it is available
for interpretation by the rule checker.

\textbf{Packet framing.} This component interprets the received symbols
according to the IEEE 802.15.4 packet definition, granting access
to header fields and payload. IEEE 802.15.4 supports several address
modes that use varying header layouts that must be supported by the
framer. This component also notifies the rule checker that a new packet
was detected via interrupts and provides a memory mapping that can
be queried to gain access to header fields and payload bytes.

\subsection{Decision Subsystem}

The decision system is triggered via interrupts by the packet framer
when a pre-defined point in the packet is reached (e.g., when the
full link layer header is available) to trigger the decision process
on whether the packet should be blocked or not. The current implementation
supports decisions based on packet content and received signal strength.
However, spoofing or replay attacks are challenging to detect based
on these features alone. Additional possibilities to classify packets,
such as more robust physical layer features (angle of arrival, device
fingerprints, etc.) are part of our future work, and are discussed
further in \prettyref{sub:Limitations}.

\textbf{Rule checker.} The rule checker classifies incoming packets
according to a pre-defined policy. It is the critical component for
real-time operation because the overall reaction time mainly depends
on its execution time. Therefore we have implemented and evaluated
two different versions: \emph{(i)} firmware code written in C and
running on the USRP2's (soft-) micro-controller, which offers runtime
reconfigurability but is comparatively slow, and \emph{(ii)} an implementation
in FPGA logic that reduces the reaction time, but the security policy
must be specified at compile time. In both implementations, the rule
checker notifies the interference subsystem with an interrupt that
a short burst of interference must presently be generated to destroy
the malicious packet.

\textbf{Rule system. }The firmware-based rule checker allows to define
content-based rules in the style of \texttt{iptables}, defining rule
chains that consist of one or more rules, each with zero or more matches
(such as source or destination address). We implemented a command
line tool (\texttt{gtables}), which generates a data structure that
can be directly interpreted by the firmware rule checker. An example
is the following rule definition with two matches (preventing the
reception of all control packets going to the broadcast address in
PAN \texttt{0x22}):

\gtables{gtables -A -m dst -{}-pan 0x22 -{}-addr 0xFFFF -m type -{}-ctrl -j DROP}

This mechanism allows to define complex access policies and deploy
them on the distributed guardians. The FPGA rule checker uses hardware
gates to compare detected packet bytes to a table of predefined values
in parallel, such that the execution time is considerably reduced.

\textbf{Administration interface.} We use the USRP2's Ethernet interface
to update the active rule set of the firmware implementation, set
operation parameters (center frequency, transmission power), and collect
system statistics, from a central point of administration. The USRP2
supports the IP protocol over Ethernet, such that all guardians can
be connected over existing networks and configured remotely by a single
administrator.

\subsection{Selective Interference Subsystem}

When a packet is classified as malicious, the guardian takes action
and prevents the reception of the packet by its protected sensor motes.

\textbf{TX waveform generator.} As described in Section~2, the efficiency
of intentional interference depends on the waveform that is transmitted,
captured in the detection threshold $\gamma_{\textrm{SIR}}$. With
a software-defined radio, arbitrary waveforms can be specified using
a sequence of I/Q samples, which completely defines a transmitted
signal. Our guardian implementation supports continuous wave (CW),
noise, and arbitrary IEEE 802.15.4-modulated symbols as interference
waveforms. 

\textbf{Transmission.} The I/Q samples are finally sent to a digital-to-analog
converter (DAC), modulated onto a carrier in the 2.4\,GHz band, and
amplified up to a maximum output power of 20\,dBm (100\,mW). External
antennas and amplifiers can be used to boost the effectively radiated
power further.

\section{Guardian Evaluation}

\label{sec:Evaluation}
\begin{figure*}
\centering \includegraphics[width=0.8\linewidth]{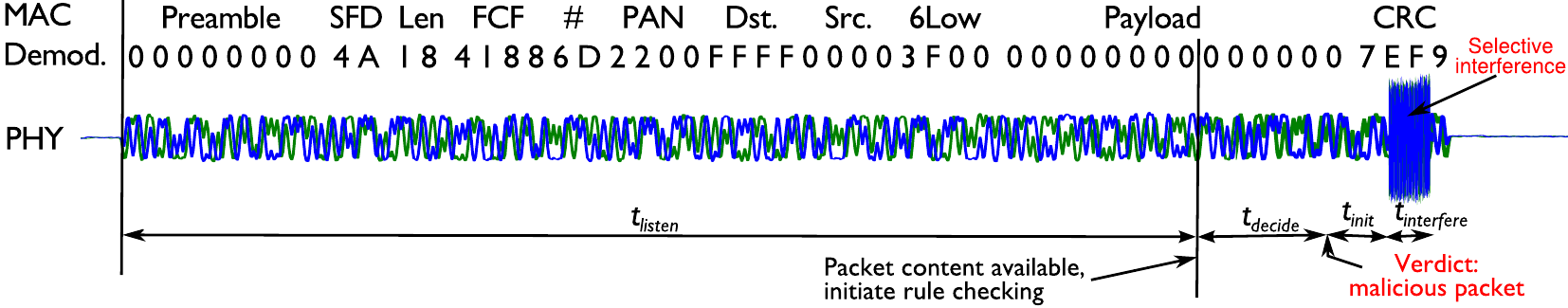}

\caption{\label{fig:Operation}Guardian in operation: first, the packet's signal
is demodulated to access its content for classification. The guardian
must wait for the payload bytes to arrive ($t_{\mathrm{listen}}$)
before the rule checker can start. When the packet is declared malicious
(after $t_{\mathrm{decide}}$), the transmission of interference is
prepared ($t_{\mathrm{init}}$), and the packet destroyed ($t_{\mathrm{interfere}}$).}
\end{figure*}
We are interested in the real-time performance characteristics of
the guardian, especially with regard to the system challenges of low-latency
and reliable packet dropping at the receiver. We evaluate how well
the different subsystems of the guardian cooperate to achieve a selective
blocking of malicious packets. The relevant time parameters of the
guardian operation (depicted in \prettyref{fig:Operation}) are described
first and then evaluated in the following subsections.
\begin{itemize}
\item The IEEE 802.15.4 receiver scans the wireless channel for relevant
signals. Once a packet is discovered, the system starts demodulating
the incoming signal, interpreting the bits as header fields according
to the standard. The rule checker starts after receiving the packet
bytes of interest, i.e., after the time duration denoted as $t_{\textrm{listen}}$.
This duration depends on how deep the packet is inspected. For example,
a typical duration for the physical and link layer headers of IEEE
802.15.4 is $t_{\textrm{listen}}=\SI{480}{\micro\second}$.
\item After the features required by the rule checker are available to the
guardian, the rule checker begins to classify the packet according
to its active rule set. For the firmware-based rule checker, this
set is traversed in sequential order. Thus, the required time (denoted
as $t_{\textrm{decide}}$) depends both on the number of rules as
well as the number of matches in each rule. For the FPGA-based checker,
several rules are checked in parallel; we measured delays below \SI{10}{\micro\second}.
\item In case the verdict of the rule checker is to destroy the packet,
the hardware initiates the interference process by preparing a waveform
and starting its transmission. We measured this delay ($t_{\textrm{init}}$)
to be small (2--\SI{3}{\micro\second}) and deterministic, so that
it is not a limiting factor in the reaction time performance.
\item Finally, an interfering signal is applied to the wireless channel,
with a duration of $t_{\textrm{interfere}}$. The signal must be sufficiently
long to reliably destroy the packet at the receiver, but should also
be as short as possible to limit the interference with co-existing
networks.
\item The overall system reaction time is denoted as $t_{\textrm{react}}$,
which is defined as the time from the start of classification to the
end of the interference, i.e., $t_{\textrm{react}}=t_{\textrm{decide}}+t_{\textrm{init}}+t_{\textrm{interfere}}$.
\end{itemize}
We proceed by first evaluating each system component described in
the previous section to determine its time parameters, and then evaluate
the overall system performance.

\subsection{Detection Accuracy}

\label{sub:Detection-Accuracy}
\begin{figure*}[t]
\centering \subfloat[\label{fig:SystemPhoto}The guardian system under evaluation in one
of our experimental settings.]{\includegraphics[width=0.3\linewidth]{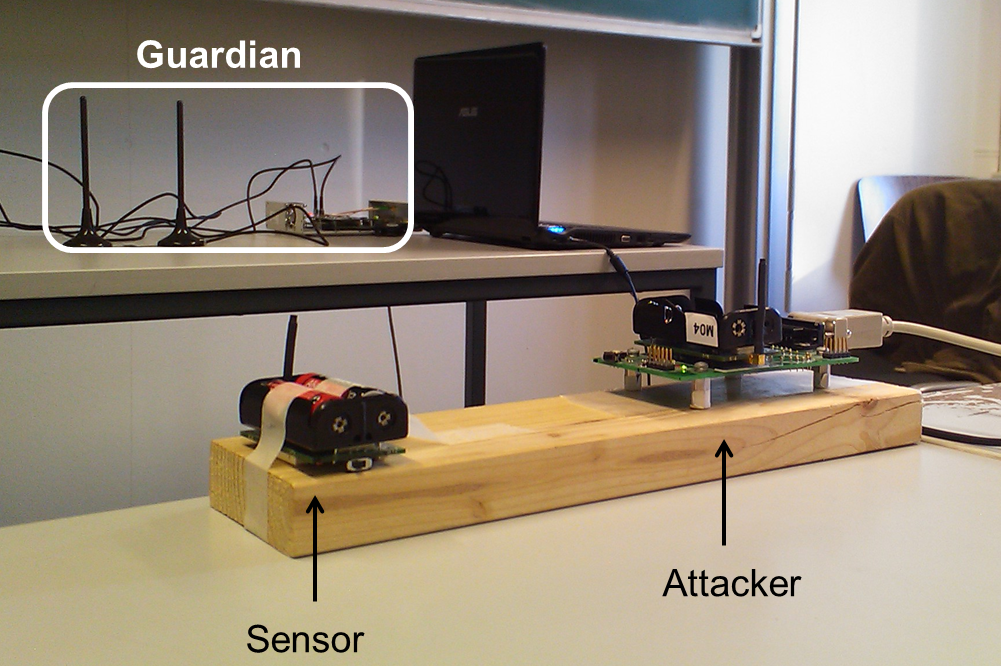}

}\hfill{}\subfloat[\label{fig:DetectPerf}Receiver sensitivity, detection performance:
packet reception ratio with increasing distance. The guardian shows
a better sensitivity, allowing to protect sensor motes from a distance.]{\includegraphics[width=0.3\linewidth]{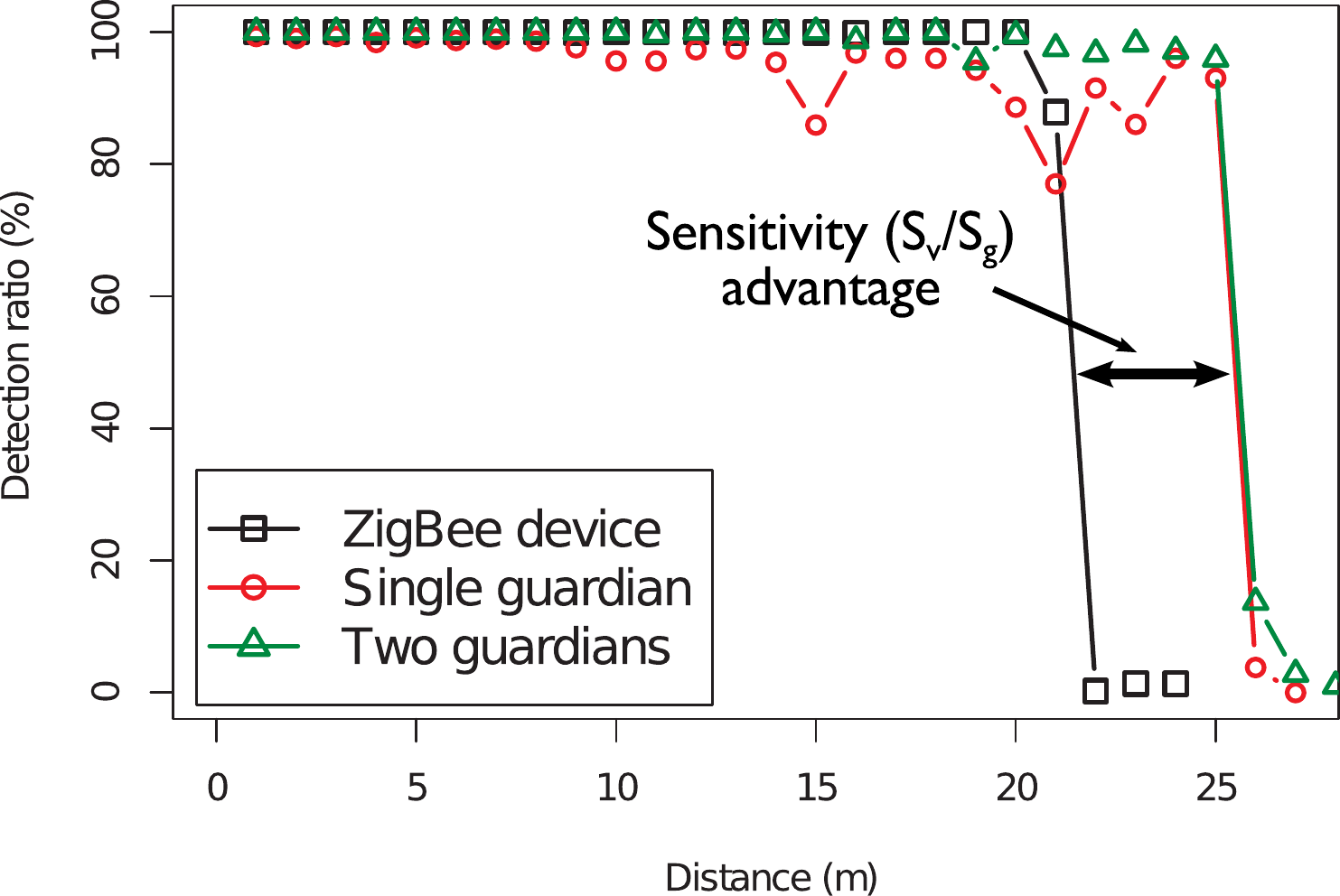}}\hfill{}\subfloat[\label{fig:RuleEvalDelay}Reaction delay for two rule checker implementations:
firmware with different rule configurations (flexible but slow) and
FPGA (static rules but fast).]{\hspace*{\fill}\includegraphics[width=0.3\linewidth]{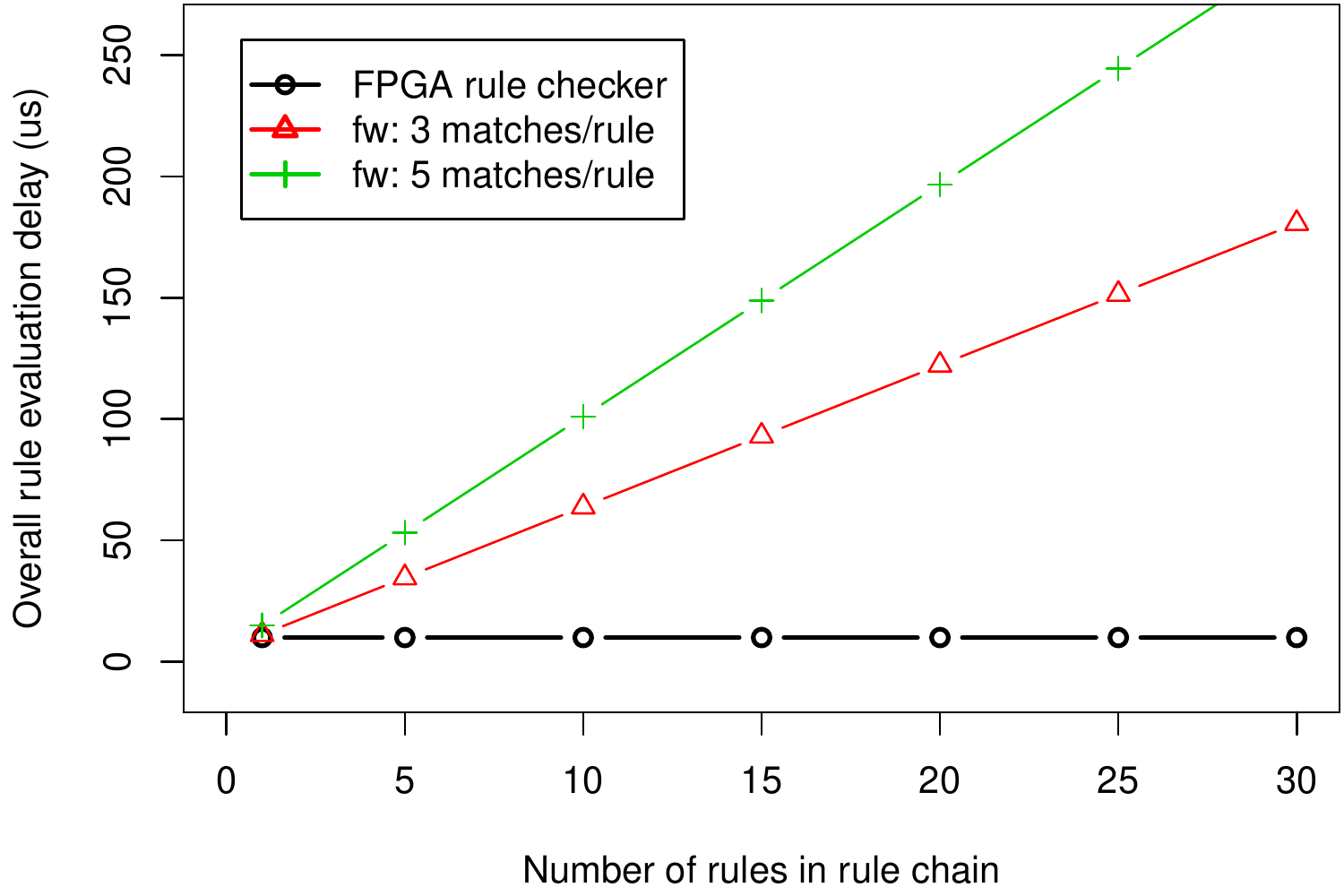}\hspace*{\fill}}

\caption{Guardian performance evaluation: experimental setting, detection performance,
and rule checking delay.}
\end{figure*}
This part of the evaluation is concerned with the speed and precision
of the packet detection subsystem. A correct operation is crucial
because only detected packets are classified (and destroyed if necessary).
We evaluate the guardian's ability to detect packets correctly while
varying the distance between attacker and guardian ($d_{ag}$). Due
to path loss, the signal strength gradually reduces, making it harder
to distinguish signal from noise. As reference, we compare the performance
of a standard-compliant MICAz receiver in this experiment to observe
if the sensitivity of the guardian is better than that of the motes
in the protected network.

\textbf{Experimental methodology:} The receiver under test (either
guardian or MICAz mote) is placed at a fixed position at one end of
the experimentation area, a 3\,m wide hallway. This indoor scenario
is a challenging test for the receiver because multipath effects and
inter-symbol-interference (ISI) increase the difficulty to detect
signals correctly. The guardian runs on a USRP2 with an XCVR2450 trans\-cei\-ver
and omni-directional antennas (3\,dBi gain). The attacker is a MICAz
mote with its default antenna, using its maximum output power (0\,dBm)
to transmit 100 packets/s (which contain 48 symbols and have a duration
of \SI{768}{\micro\second}) for 10\,s.

Starting from a 1\,m attacker--receiver distance, we then move the
attacker to a set of measurement positions with distances $d_{ag}=1,\ldots,30$\,m.
For each position, the attacker transmits 1000 broadcast packets.
The receiver (guardian or MICAz) counts the successfully detected
packets (confirmed by a CRC check) for the packet reception ratio.

\textbf{Results:} The detection results are shown in \prettyref{fig:DetectPerf}.
The COTS receiver performs well in this experiment, with a detection
radius of approximately 20 meters. With 12 out of 20000 packets in
this range, only a limited number of packets was not received successfully.
Our FPGA-based receiver implementation has an increased reception
radius (up to 25\,m), indicating a higher sensitivity compared to
the MICAz mote. However, the performance for longer distances deteriorates
slightly because the receiver is sensitive to multipath fading effects,
causing a fraction of packets to be missed in some locations. We mitigate
this effect by using a second guardian. In this setup, the number
of missed packets is again reduced. Overall, the system fulfills the
goal of a better receiver sensitivity discussed in Section~2, such
that the guardians are suitable to protect sensor nodes from a distance.

\subsection{Reaction Delay}

\label{sub:ReactionTime}
\begin{table}
\centering %
\begin{tabular}{ccccc}
Blocking rule & Byte offset & Max.~$t_{\textrm{react}}$ & Firmware & FPGA\tabularnewline
\hline 
Start-of-Frame Delim.  & 5 & \SI{864}{\micro\second} & $\surd$ & $\surd$\tabularnewline
Frame Control Field  & 7--8 & \SI{768}{\micro\second} & $\surd$ & $\surd$\tabularnewline
Source Address  & 14--15 & \SI{544}{\micro\second} & $\surd$ & $\surd$\tabularnewline
Payload byte \#16 & 27 & \SI{160}{\micro\second} & $\surd$ & $\surd$\tabularnewline
Last payload byte  & 30 & \SI{64}{\micro\second} & $\times$ & $\surd$\tabularnewline
\end{tabular}

\caption{\label{tab:PacketInpectTimes}Impact of overall reaction delay on
feasible blocking rules. The symbol $\surd$ indicates that the guardian
can use the respective blocking rule and still destroy the packet.}
\end{table}
By reaction delay $t_{\textrm{react}}$, we refer to the time from
receiving the content bytes of interest until the interference is
finished. This delay affects how deep the guardian can look into a
packet, because the interference must overlap with it. It is mainly
dependent on the execution time $t_{\textrm{decide}}$ of the guardian
rule checker to decide whether the packet should be destroyed. An
overly slow decision process may shift the selective interference
behind the end of the packet. We put both firmware and FPGA-based
rule checker to the test.

\textbf{Experimental methodology.} The measurements are taken using
the FPGA's 100\,MHz internal clock to record timestamps, which allows
us to reach a timing precision of 10\,ns. The rule evaluation is
timed from the instant an hardware interrupt signals the detection
of an IEEE 802.15.4 header by the framer to the instant the rule system
returns a verdict on how to treat the packet. In the firmware-based
rule system, this is the return of the C function call that activates
the rule checker. For the FPGA-based rule checker, the timestamp is
taken when the rule checker interrupt to start the interference arrives
in the firmware. 

We vary the number of rules in the chain, and the number of matches
in each rule (each match is a C function that reads data from the
framer and compares these values to constants stored in the chain).
The rules are chosen such that none of them matches the packet, hence
we measure the worst-case run time where the chain is traversed completely.
As the rules are evaluated in parallel in the FPGA-based implementation,
increasing the rule set size affects the FPGA resource usage and not
the timing.

\textbf{Firmware-based rule checker results.} The compound execution
times depicted in \prettyref{fig:RuleEvalDelay} show that the reaction
delay is depending on the rule set used. The reason is that our implementation
follows the general design of \texttt{iptables} using linked lists
with variable size, such that a small overhead occurs for the evaluation
of each rule chain, rule, and the matches contained in a rule to traverse
the list. As the used micro-controller only supports a single thread,
the required execution time is deterministic but increasing with each
rule.

To break the delays into components, we analyzed the rule checker
implementation in depth. First, a constant time of $\SI{4.03}{\micro\second}$
is needed to enter the interrupt handler, jump into the rule checker
and back, and trigger the interference process. This cost is independent
of the chain's contents and is paid for each detected packet. To evaluate
one rule in the chain, the guardian needs $\SI{0.26}{\micro\second}$
for evaluation (mainly the time to traverse the chain). Evaluating
the time needed for individual rule, each match needs $\SI{0.34}{\micro\second}$
to start the execution of the associated test function. The overall
running time of a match function depends on the logic of the match
itself. Considering the address match, a match with representative
complexity, the execution takes $\SI{1.86}{\micro\second}$ before
the function returns. All matches in a rule are checked sequentially,
such that the execution times add up. For example, the execution of
one rule with 3 matches accounts to $\SI{5.58}{\micro\second}$. So,
when considering a chain with 20 rules of 3 matches each, the guardian
requires $\SI{116}{\micro\second}$ or approximately 4 payload bytes
to react in the worst case (when all rules are traversed).

\textbf{FPGA-based rule checker results.} To allow more deterministic
decision delays and to support a deeper look into the packet, we also
implemented a rule checker implemented in FPGA logic that is less
flexible but provides faster reaction times. Using this approach we
are able to cut the latency down to a limited number of FPGA clock
cycles, i.e., below one microsecond up to \SI{10}{\micro\second}.
This enables us to achieve an overall reaction time of $t_{\textrm{react}}=\SI{39}{\micro\second}$,
even with complex rule sets. This means that the guardian can base
its decision on the complete payload and still hit the CRC bytes at
the end of the packet to cause a packet drop.

\subsection{Interference Waveforms and Duration}

\label{sub:JamDuration}
\begin{figure*}
\centering \subfloat[\label{fig:JamWaveforms}Comparison of the packet reception ratio
for three interference waveforms: efficient waveforms increase the
detection threshold $\gamma_{\textrm{SIR}}$ and thus require less
transmit power for blocking.]{\includegraphics[width=0.3\linewidth]{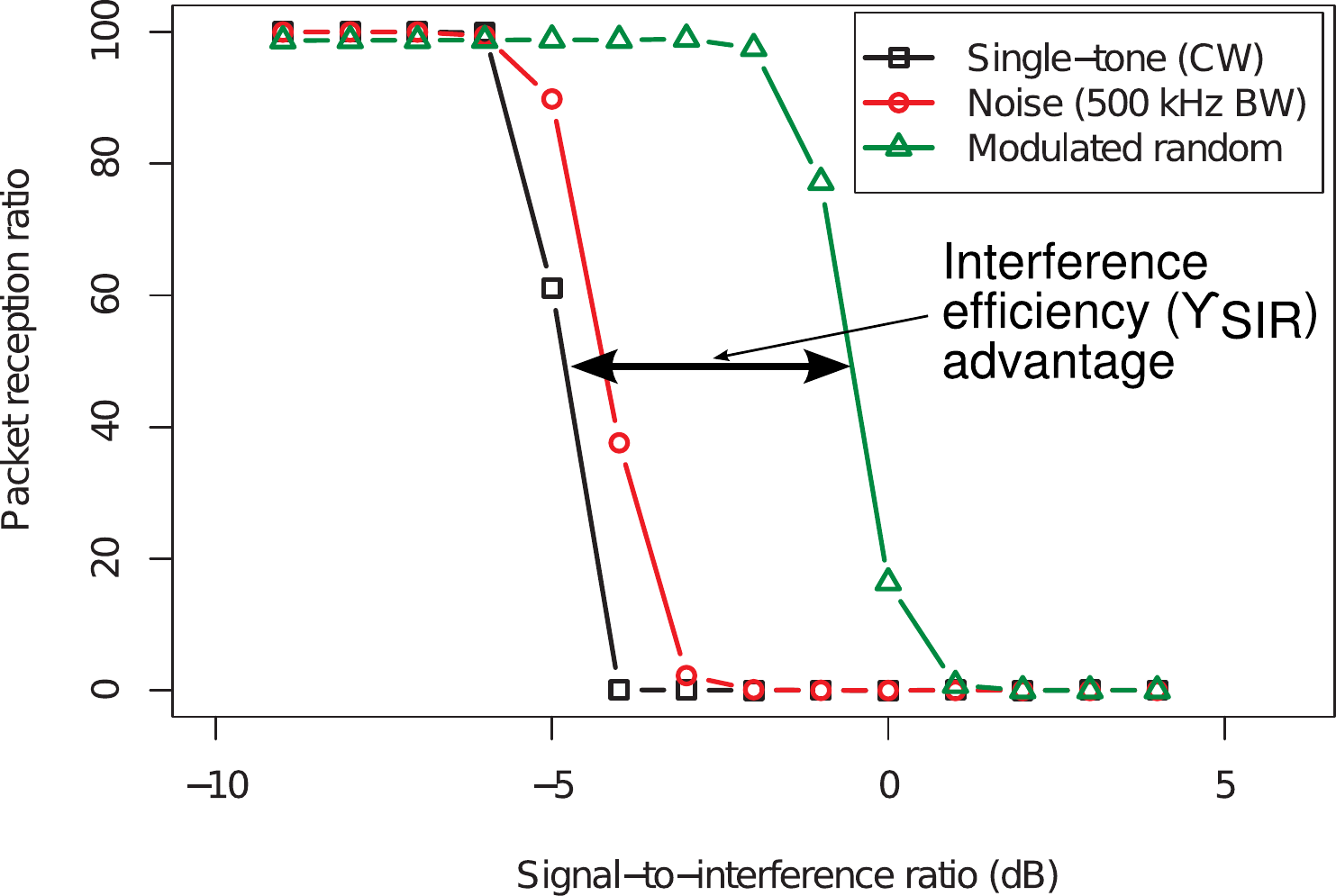}}\hspace*{\fill}\subfloat[\label{fig:JamDuration}Minimum interference duration for IEEE 802.15.4
radios, interfering with \SI{26}{\micro\second} of a packet transmission
is sufficient to trigger packet drops at the receiver reliably.]{\includegraphics[width=0.3\linewidth]{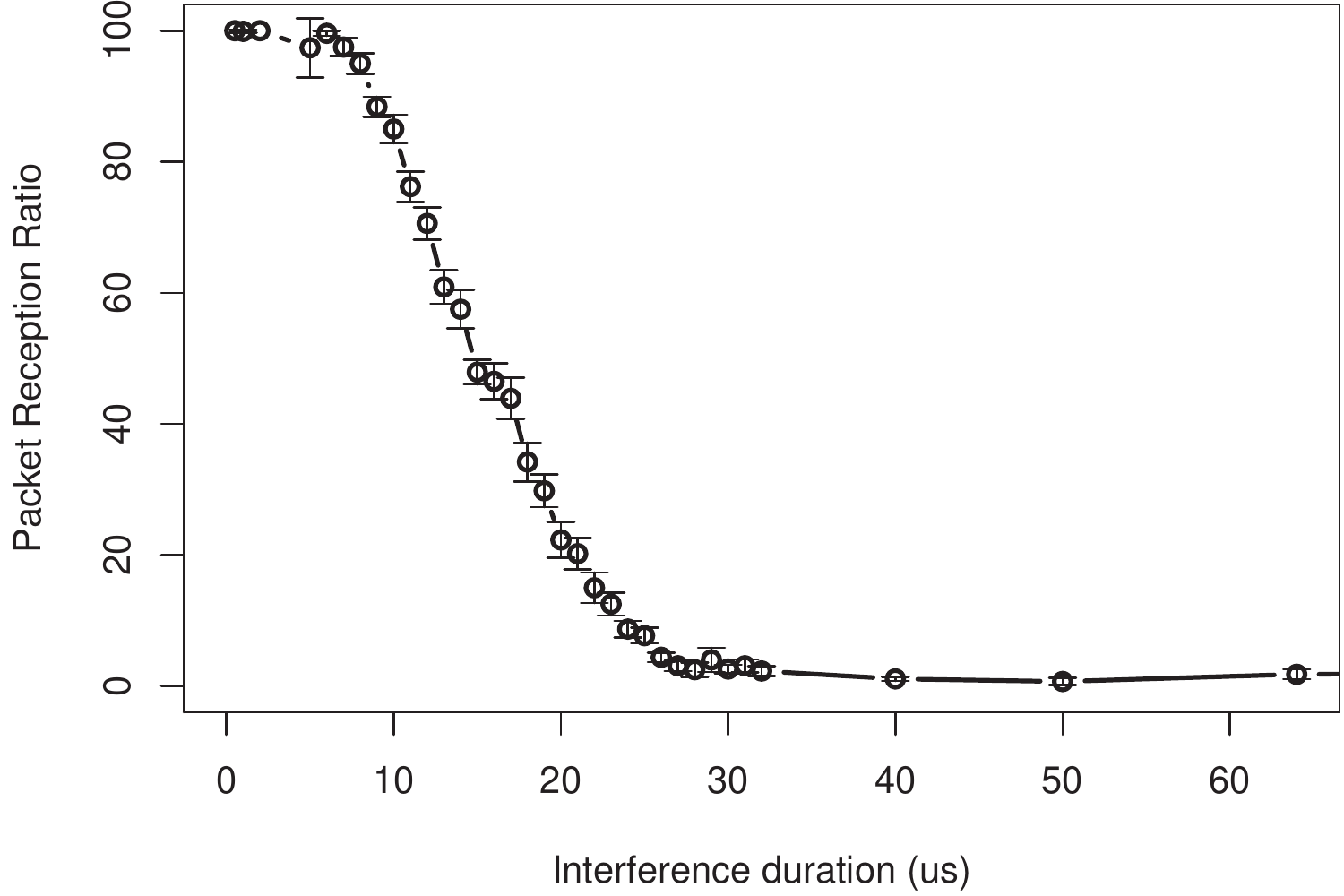}}\hspace*{\fill}\subfloat[\label{fig:Effectiveness-Jam-15}The system's overall protection performance
with close-proximity attacker (the attack always succeeds without
guardians): detecting, classifying, and destroying the attacker's
packets.]{\includegraphics[width=0.3\linewidth]{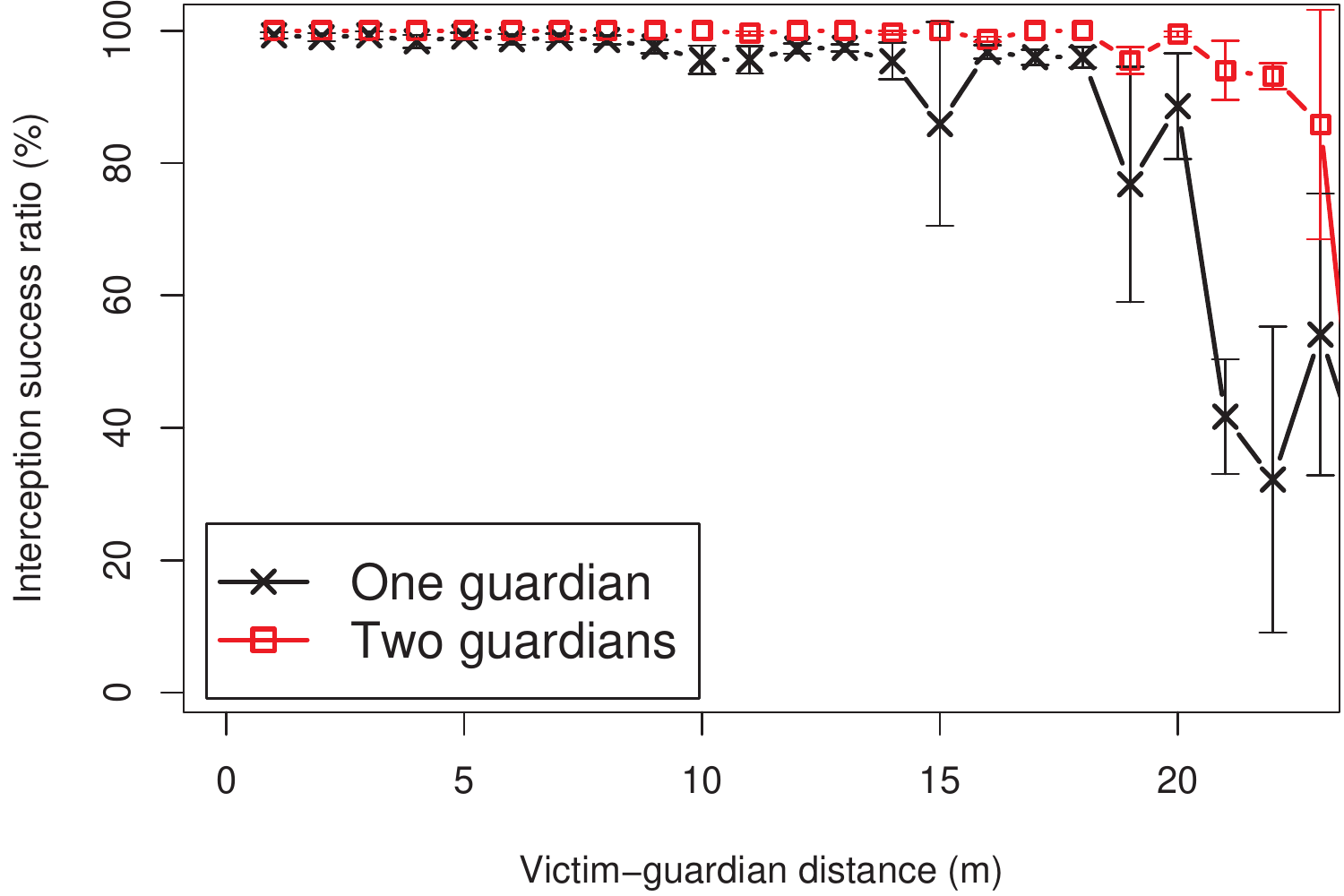}}

\caption{\label{fig:Eval-Detection}Performance evaluation: interference waveforms,
minimum interference duration, and overall system performance.}
\end{figure*}
We are interested in the optimal waveform that offers a high chance
to intercept a packet (by providing a beneficial~$\gamma_{\textrm{SIR}}$),
and the minimum interference duration that introduces bit errors into
the packet with high probability.

\textbf{Interference waveform.} We tested three candidate waveforms
for their ability to prevent a packet reception: a continuous wave
(CW) positioned at the center frequency of the channel, white noise
with a bandwidth of 500\,kHz around the center frequency, and random
symbols spread and modulated as specified in IEEE 802.15.4. We evaluated
their performance according to the required transmit power to achieve
a packet reception ratio of 0\,\%. The results are shown in \prettyref{fig:JamWaveforms}:
the continuous wave offers the most energy-efficient way to interfere
with reception against the sensor platform we use, with the additional
benefits that it is easy to generate on the SDR and that co-existing
technologies such as WLAN filter out this type of interference relatively
well, hence their operation is not disturbed by the guardian.

\textbf{Minimum interference duration.} Using the CW waveform, we
evaluated the minimum interference duration on the channel to reliably
destroy a packet. Guardian and victim are placed in close vicinity
such that the guardian's interference is stronger than the signal
of each packet (see \prettyref{fig:SystemPhoto}). The results in
\prettyref{fig:JamDuration} show that an interference duration of
\SI{26}{\micro\second} is sufficient to destroy a packet. This result
has two implications: first, the energy cost of the attacker is high;
while the attacker must transmit a complete packet to be successful
(e.g., \SI{1024}{\micro\second} for a 32\,byte packet), the guardian
invests 40 times less energy to successfully prevent the reception.
Second, the interference duty cycle of the guardian is very low, minimizing
the effect on co-existing networks. In the experiments in the next
section (100 attack packets/s), the guardian only transmits for less
than one millisecond each second. We discuss the issue of interference
further in \prettyref{sub:Legal}. The presented experiments are based
on preliminary results published in \cite{ReactThreat}.

\subsection{Protection Performance}

\label{sub:JamPerf}The most important aspect of our system is to
effectively block malicious packets at the sensor motes. We consider
the performance of the overall system here, i.e., both detection and
selective interference are used in combination.

\textbf{Experimental methodology.} We place one or two guardian(s)
at one end of the hallway, with the same measurement positions as
in the detection experiments, i.e., $d_{ag}=1,\ldots,30$\,m. Attacker
and victim (both MICAz) are moved together with a constant distance
of one meter ($d_{av}=1$\,m), providing the attacker with excellent
conditions. For each position, the attacker sends 100 packets/s back-to-back
with 0\,dBm output power. Under this attack rate, the guardian must
also prove that the transmission operation does not affect its subsequent
detection performance and that its turnaround time is small enough
to support high frequency attacks.

We repeat this attack 10 times for each distance, slightly moving
the attacker each time, to lower the effects of multipath fading on
the experimental results. We use the firmware-based rule system; the
rule used is to search for all broadcast packets originating in the
current PAN:

\gtables{gtables -A -m dst -{}-addr 0xFFFF -{}-pan 0x22 -j DROP}

All packets sent by the attacker match this specification, thus the
guardian has to interfere with all attacker packets.

\textbf{Results.} A single guardian successfully prevents a packet
reception in 98--99\,\% for distances up to 15\,m, despite the close
proximity of attacker and victim (see \prettyref{fig:Effectiveness-Jam-15}).
The outcome of this experiment is similar to the detection experiment,
suggesting that the system performance is mainly influenced by the
guardian's receiver implementation. Similar to the detection case,
two guardians located at the same position effectively counter the
issue of 1--2\,\% packet misses, and achieve a combined 99.9\,\%
protection rate up 18\,m. The two guardians only missed 19 out of
18000 packets up to this distance.

\subsection{Summary}

\begin{table}
\centering %
\begin{tabular}{cccc}
Parameter & Description & Firmware & FPGA\tabularnewline
\hline 
$t_{\textrm{decide}}$ & Rule checker execution time (var.) & \SI{116}{\micro\second} & \SI{10}{\micro\second}\tabularnewline
$t_{\textrm{init}}$ & Duration decision--start transmitting & \SI{3}{\micro\second} & \SI{3}{\micro\second}\tabularnewline
$t_{\textrm{interfere}}$ & Interference duration for packet drop & \SI{26}{\micro\second} & \SI{26}{\micro\second}\tabularnewline
\hline 
\bfseries$t_{\textrm{react}}$ & \bfseries Overall reaction delay & \bfseries\SI{145}{\micro\second} & \bfseries\SI{39}{\micro\second}\tabularnewline
\end{tabular}

\caption{\label{tab:TimeConstants}Time parameters of our guardian implementation,
indicating that packets can still be destroyed after observing a large
part of their contents.}
\end{table}

The guardian implementation allows real-time detection of malicious
packets with a high accuracy during their transmission and a reliable
destruction before the packet may arrive at a receiver. The results
are summarized in Tables~\ref{tab:PacketInpectTimes} and \ref{tab:TimeConstants}.
The system is able to classify and destroy 99.9\,\% of the packets,
even if the classification depends on the last byte in the payload,
because the reaction time of the system (\SI{39}{\micro\second})
is shorter than the duration of the CRC field (\SI{64}{\micro\second}).

\section{Applications}

\label{sec:Apps}This section demonstrates the ability of our guardian
system to control wireless channel access in order to protect IEEE
802.15.4 wireless sensor networks from message injection attacks.
To evaluate the effectiveness of our guardian system, we set up different
use cases in our lab and implement real-world attacks using open-source
attack tools from the KillerBee suite \cite{Killerbee}.

\subsection{Experimental Setup}

\begin{figure}
\centering \includegraphics[width=0.8\columnwidth]{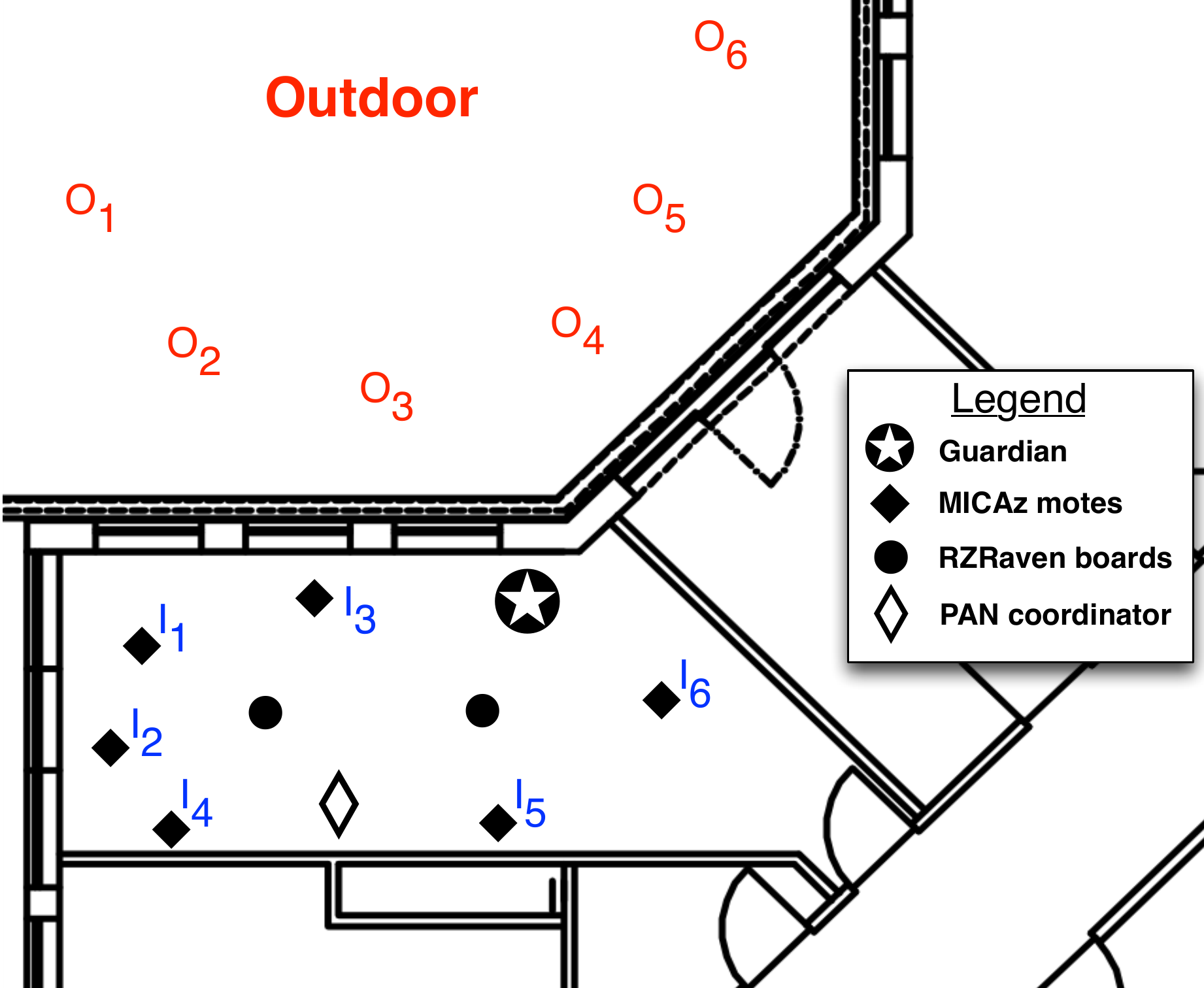}

\caption{\label{fig:Experimental-setup.}Floor plan and experimental setup.
The experiments are conducted in a university lab on the second floor,
with several WLAN access points working in the same frequency band.
Outside positions are located on the roof of the building.}
\end{figure}

\textbf{Radio network environment.} The experiments are conducted
in a fairly typical office environment, located on the second floor
of our university building. The IEEE 802.15.4\slash{}ZigBee network
consists of heterogeneous devices operating on the 2.4\,GHz frequency
band. The WSN includes the following devices (a floor plan with the
positions of deployed devices is shown in \prettyref{fig:Experimental-setup.}):
\emph{(i)} one PAN coordinator using the Atmel AVR Wireless Service
Suite under Windows XP on a 2.0\,GHz machine with an Atmel RZ USB
transceiver, \emph{(ii)} two Atmel AVR RAVEN boards equipped with
a picoPower AVR processor, and \emph{(iii)} six Crossbow MICAz devices
running TinyOS.

\textbf{Attacker configuration.} The attacker is equipped with an
802.15.4-compliant ZigBee transceiver flashed with the KillerBee firmware.
It can launch a variety of automated attack vectors against ZigBee
networks. Particularly interesting are attacks which abuse legitimate
protocol operations like launching resource-depletion (DoS) attacks
or the intercepting and manipulating over-the-air (OTA) configurations
(including firmware upgrades and key exchanges).

\textbf{Guardian configuration.} Depending on the respective scenario,
we change the type of antenna (either omni-directional or sector)
and its orientation (inside\slash{}outside). We use COTS 2.4\,GHz
antennas, with standard specifications.%
\footnote{Detailed antenna specifications of the antennas used: 
\begin{itemize}
\item Omni-directional antenna: length: 12\,cm, gain: 5\,dBi.
\item Directional (sector) antenna: length: 16\,cm, beam width horizontal:
$69.6^{\circ}$, vertical: $64^{\circ}$, max.~gain: 8\,dBi. A data
sheet is available at \url{http://www.wimo.com/download/18560_8.pdf}\end{itemize}
}

\subsection{Protecting from Unauthorized Messages\label{sub:App--KillerBee}}

\begin{figure}
\begin{centering}
\centering \includegraphics[clip,width=0.9\columnwidth]{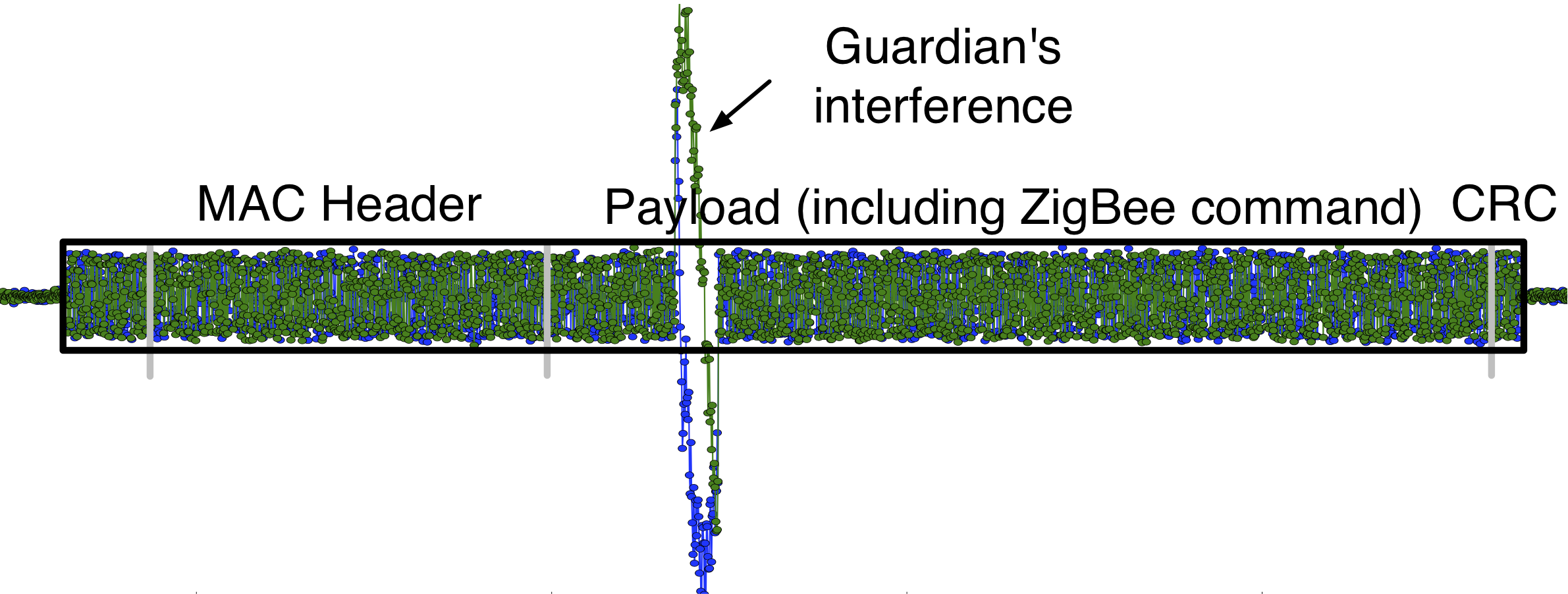}
\par\end{centering}

\centering{}\caption{\label{fig:Deep-packet-inspection}Preventing an unprotected OTA firmware
upgrade with selective interference (measured physical layer baseband
trace).}
\end{figure}

\subsubsection{Controlling Over-the-Air Updates}

In this experiment, an attacker tries to modify the firmware of a
ZigBee device with an Over-The-Air (OTA) configuration update. The
off-the-shelf OTA update functionality of the RZ RAVEN boards does
not require any authentication. The only requirement is that the node
initiating the update is associated to the PAN. The attack is implemented
using KillerBee: with the \texttt{zbreplay} tool and a spoofed address
of the PAN coordinator, an arbitrary firmware is uploaded to the RZ
RAVEN devices. To avoid such vulnerabilities, the guardian blacklists
all OTA configuration traffic. In order to enforce such a policy,
the guardian have to perform a payload inspection because the information
about packets carrying OTA configuration commands can only be found
in higher layers of the ZigBee protocol stack. More specifically,
it checks for data frames with a certain network layer control field
(\texttt{0x0008}), and a certain ZigBee application support layer
command type (\texttt{0x01}) in the payload of the packet, resulting
in the following rule:

\gtables{gtables -A -m dst -{}-pan 0xACAC -m nw\_ctrl 0x0008 -m asl\_cmd 0x01 -j DROP}

\prettyref{fig:Deep-packet-inspection} shows the baseband trace of
an OTA firmware upgrade packet while it is being blocked by the guardian.
A short, yet precise selective interfering signal hits the middle
of the packet and renders it unusable. Hence, the attacker is no longer
able to modify the firmware of the RZ RAVEN nodes in the vicinity.

\subsubsection{Mitigating Resource-Depletion Attacks}

\label{sub:App--ResourceDepletion}
\begin{figure}
\begin{centering}
\centering \includegraphics[clip,width=0.9\columnwidth]{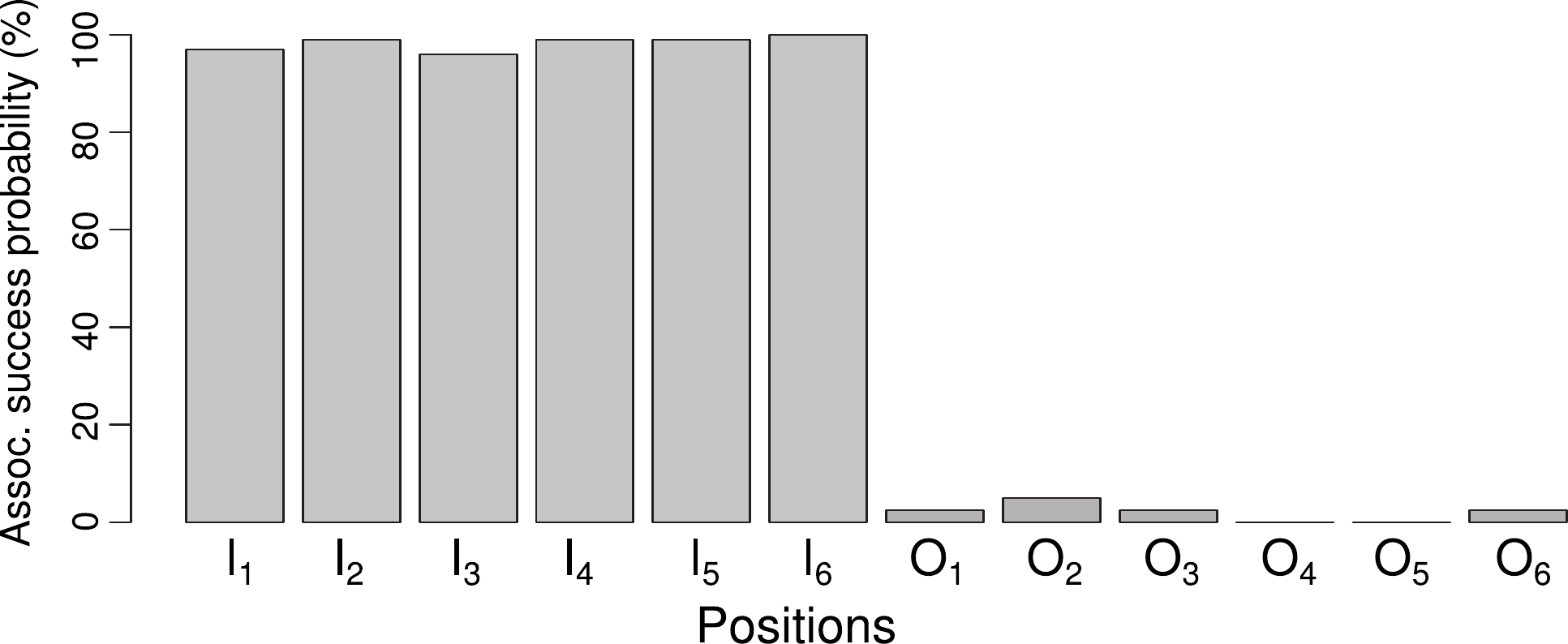}
\par\end{centering}

\centering{}\caption{\label{fig:AssociationFlooding}Resource-depletion attack by flooding
the network with association requests. Outside attackers are effectively
blocked, while indoor operations are not affected.}
\end{figure}
Next, the KillerBee toolbox is used to launch a resource depletion
attack on the ZigBee devices. The attack is executed by KillerBee's
\texttt{zbflood} tool, which floods the PAN coordinator with association
requests (IEEE 802.15.4 control frames). Under normal conditions,
we verified that this causes our PAN coordinator to crash repeatably.
To make the coordinator operational again, the node must be reset
manually after each crash.

To mitigate this attack, assuming that the attacker is located outside
the room, we deploy the following guardian rule:

\gtables{gtables -A -m type -{}-control -m dst -{}-pan 0xACAC -m RSS -{}-above -80 -j DROP}

The guardian's policy is to destroy all association requests (IEEE
802.15.4 control frame type \texttt{0xC823}) sent to the network ID
\texttt{0xACAC}, and transmitted from outside. The idea of this policy
is to allow associations from inside the room, but not from outside
(assuming access to the room is physically secured). To that end,
we define outside as an area covered by the sector antenna and a signal
threshold exceeding $-80$\,dBm.%
\footnote{Admittedly, this definition of outside is not particularly robust,
but can be improved by more sophisticated signal features (see the
discussion in \prettyref{sub:PHYFeatures}). The assumption is the
guardians are placed at the border between inside and outside and
the attacker's effective power must be at least $-80$\,dBm at the
guardian's position to reach the sensor motes located behind it.%
} The guardian's response is to generate interference using its omni-directional
antenna. We are interested in the successful association request rates
from inside (sent by legitimate sensors) vs.~outside positions (sent
by KillerBee). The tested positions are marked in \prettyref{fig:Experimental-setup.}
as $I_{1},\dots,I_{6}$ for inside, and $O_{1},\dots,O_{6}$ for outside
positions. As can be observed in \prettyref{fig:AssociationFlooding},
the guardian allows association requests from inside positions with
a probability of over 95\,\%. However, for senders closer to the
sector antenna, the guardian occasionally blocks ($\rightarrow$ positions
$I_{1}$ and $I_{3}$). This is an artifact of the simple location
detection scheme used. Nevertheless, the attacker's flooding rate
from outdoor positions is severely limited and this rate is no longer
sufficient to make the PAN coordinator crash from any outdoor position.

\subsection{Instant Node Revocation\label{sub:App--Revocation}}

The next attack scenario considers node capture and replication attacks
in WSNs \cite{TMC_NodeReplica,TMC_ReplicDetect}. While this problem
is mainly treated as a key management issue in the literature \cite{OnRevoc,NodeReplic},
we show that compromised sensor motes can also be removed on the physical
layer once they are identified. We refer to this as \emph{instant}
revocation because once the blocking rules are committed to the guardian,
the motes' channel access is instantaneously blocked and they are
thus disconnected from the network; it is not necessary to reliably
distribute a revocation command in the (possibly Byzantine) network.
From the guardian's perspective, the channel control policy is to
detect packets from revoked nodes by their source addresses (sensors
\texttt{0x1111} $(I_{1})$, \texttt{0x1112} $(I_{3})$, \texttt{0x1115}
$(I_{5})$) and network ID (\texttt{0xACAC}) and to destroy those
packets: \vspace{3mm}

\gtables{gtables -A -m src -{}-addr 0x1111 -{}-pan 0xACAC -j DROP\\
gtables -A -m src -{}-addr 0x1112 -{}-pan 0xACAC -j DROP\\
gtables -A -m src -{}-addr 0x1115 -{}-pan 0xACAC -j DROP}\vspace{4mm}

\begin{figure}
\begin{centering}
\centering \includegraphics[clip,width=0.8\columnwidth]{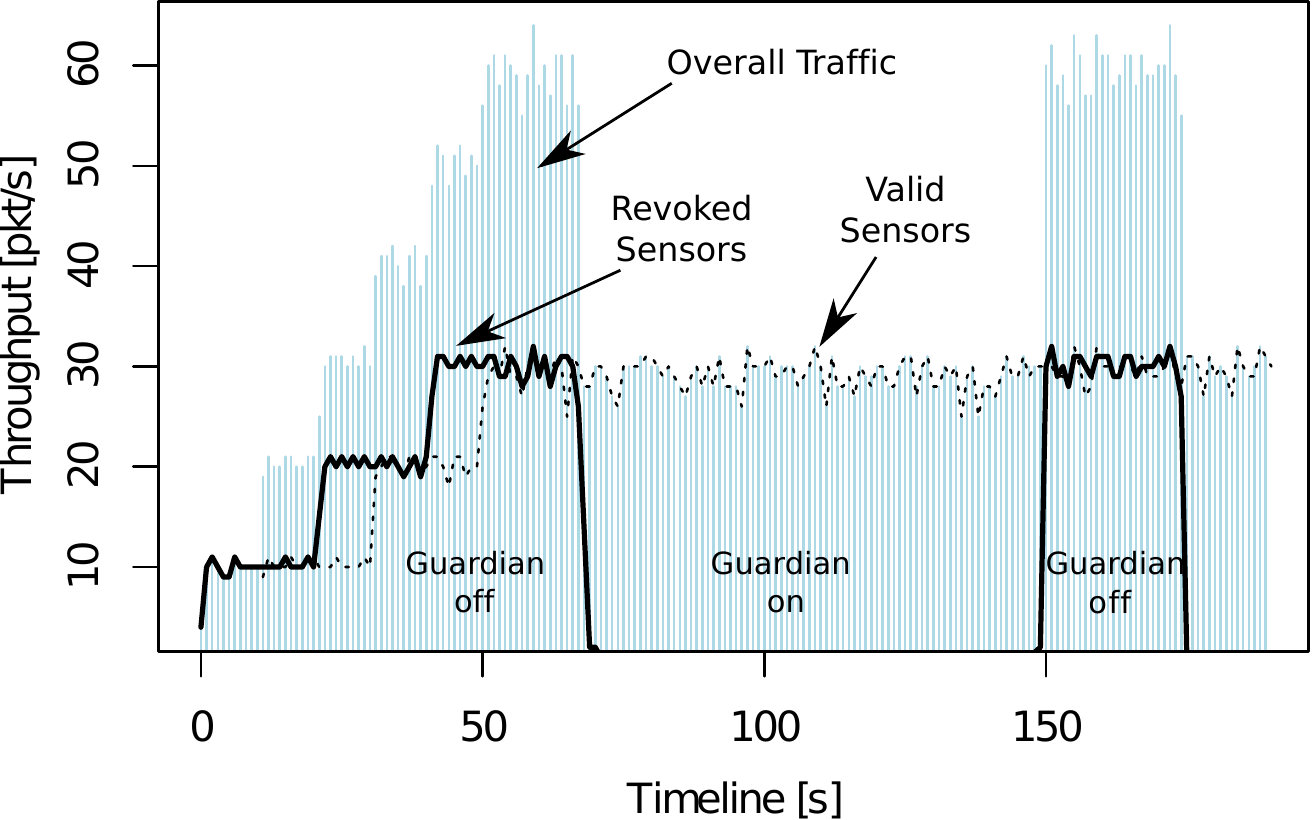}
\par\end{centering}

\centering{}\caption{\label{fig:Rapid-node-revocation}Instant node revocation: 6 MICAz
motes are consecutively turned on and start transmitting. After 70
seconds, the guardian is turned on to selectively block traffic transmitted
from revoked devices (3 motes).}
\end{figure}
In this experiment, six MICAz motes consecutively start transmitting
with 10 packets/s. After 70 seconds, three nodes are revoked for 90
seconds, then allowed again for 20 seconds, and finally revoked for
the rest of the experiment. We are interested in packets from revoked
nodes able to reach the network (false negatives) and the impact of
the guardian on the legitimate traffic (false positives). 

The results are shown in \prettyref{fig:Rapid-node-revocation}. The
stepwise traffic increase is due to the consecutive start of the transmissions.
The black solid line is the cumulative traffic of the nodes to be
revoked, the dashed line shows the traffic of legitimate nodes, and
the overall traffic is depicted by the bars in the background. As
can be seen, the guardian immediately reacts by completely blocking
the traffic from the revoked nodes. During the revocation phases,
the amount of legitimate traffic equals the overall traffic, so there
are no false positives. The number of false negatives is one packet
at the beginning and at the end of revocation phases (due to the transition
of the guardian's rule re-configuration).

\section{Discussion}

\label{sec:Discussion}In this section, we provide a discussion on
how to lift limitations of our guardian implementation, some non-technical
aspects of its use, and future work opportunities.

\subsection{Lifting Limitations}

\label{sub:Limitations}We first discuss functional extensions to
support rules based on physical features and multi-channel operation
to enable additional applications.

\subsubsection{Going Beyond Content}

\label{sub:PHYFeatures}The current system implementation supports
classification based on the content of the packet, such that certain
types of attacks (spoofing and replay attacks) are hard to single
out. However, the packet content is not the only feature extractable
from incoming packets: the physical characteristics of a signal are
influenced by RF propagation phenomena and transmitter characteristics.
We already exploited such signal ``meta-data'' in the DoS application
scenario (\prettyref{sub:App--ResourceDepletion}) by giving the rule
system access to the received signal strength, thwarting a DoS attack
that used replayed packets. Still there is a wide range of additional
physical features that can increase the attack detection options of
our guardian system. We briefly introduce a selection of features
considered in the literature.

\textbf{Energy-based positioning:} Several methods to infer a sender's
position are proposed in the literature. The use of power level information
such as RSS is used to position devices precisely \cite{infoc_RADAR},
even to localize them in large scenarios \cite{mobc_LargeScaleLocalization}.
This enables location-aware applications such as \emph{geo-fencing}
\cite{perv:GeoFencing}. The guardians could then benefit from rules
that use the sender location in their decision and detect spoofing
attacks \cite{DetectSpoofRSS}.

\textbf{Direction-based positioning:} With antenna configurations
such as directional antennas or antenna arrays, guardians can also
gain capabilities of position-based classification; a simple example
use is shown in \prettyref{sub:App--ResourceDepletion}. These methods
use the direction of arrival to infer the position of a transmitter.
Angle of arrival information has already been shown to be valuable
for securing WLANs \cite{SecureAngle}.

\textbf{Link signatures:} A more complex method using physical layer
information to position devices is the use of link signatures \cite{mobc_temporal_link_signatures}.
This method is able to discern the position of two devices with a
large probability using spectral information and may be used to prevent
spoofing and replay attacks.

\textbf{Device identification:} This analysis method enables to uniquely
identify transmitting devices \cite{OnPHYID}. For example, the concept
of device fingerprints \cite{mobc:brik-device-id} uses imperfections
in the TX chain of transmitters to associate packets to the transceiver
hardware used. These features are stable and unique enough to identify
devices even from the same production line. This technique can be
used to whitelist trusted devices, blocking adversaries that cannot
mimic the physical layer behavior of the devices.%
\footnote{While spoofing can still be achieved \cite{AttackPHY}, the attacker
must increase its effort significantly.%
}

\subsubsection{Protecting Multiple Channels in Parallel}

Currently, the guardian implementation supports the protection of
a single IEEE 802.15.4 channel. However, it can be extended to support
the monitoring of multiple channels in parallel, by replicating the
relevant subsystems in the FPGA for each channel. This only alters
the amount of resources used, not the timing behavior because of the
inherent parallelism of the FPGA architecture. Recent results show
that this approach is feasible by demonstrating that four IEEE 802.15.4
channels can be received concurrently on the USRP2 platform \cite{MultiChannel}.

\subsection{Non-technical Aspects}

\label{sub:Legal}

\textbf{Legal Aspects.} The intentional generation of interference
may raise the concern whether our guardians can be operated legally.
In general, this question is not simple to answer because the rules
governing spectrum access vary across countries and frequency bands.
The U.S.~regulations \cite{FCC15} mandate in $§$15.5 that ``harmful
interference,'' an emission that ``obstructs or repeatedly interrupts
a radiocommunications service operating in accordance with {[}Chapter
15{]},'' is forbidden. However, as we limit the interference to adversarial
packets by analyzing and deciding on a per-packet basis, we might
argue that no ``service'' is interrupted. 

Regarding unintentional interference with co-existing networks, we
point out that the guardian accesses the channel scarcely. The guardian
emits a \SI{26}{\micro\second} signal per packet and is silent for
the rest of the time (e.g., \SI{998}{\micro\second} for 32\,byte
packets). From the view of a single channel, such a behavior is also
observed for frequency hopping systems such as Bluetooth. In fact,
Bluetooth Power Class~1 devices \cite[$§$7.2]{IEEE802.15.1-2005} use
the same transmit power (100\,mW) as our guardians and occupy a 2\,MHz
802.15.4 channel for approximately 25\,ms per second, which is comparable
to the emissions of a guardian reacting to an attacker with maximum
rate (1000 packets/s). This also shows that attacking the guardian
infrastructure itself (by deliberately triggering interference) leads
only to a limited channel occupancy. In addition, as we consider the
operation in the 2.4\,GHz band, co-existing devices such as IEEE
802.11 receivers commonly filter out the simple sinusoidal waveform
we chose. Thus, while a comprehensive discussion of the legal aspects
of RF interference is out of scope for this paper (as well as out
of our expertise), we observe that the selective and efficient operation
of the guardians effectively limits interference, and finally remark
that sometimes legislation follows technical innovation.

\textbf{Economic Aspects.} The guardians are additional infrastructure
that is deployed alongside the sensor network. While this offers several
benefits (central control, on-demand security, or the possibility
to ``patch'' legacy networks), it may raise the question of cost.
The number of guardian devices depends on the WSN deployment area
and on the desired level of protection. Thus, the cost per device
should be small; one option is to implement the system with COTS transceiver
and micro-controller chips, possibly sacrificing flexibility. On the
other hand, recent results also show that it is possible to build
cheaper ($\sim$\$\,100) and more energy-efficient SDR platforms
\cite{SDRDiet}.

\subsection{Future Work Opportunities}

There are several interesting research opportunities to extend and
refine the guardian concept. In addition to an extension with more
physical layer protocols and physical feature detectors, we briefly
discuss two promising paths.

\textbf{Optimal guardian deployment.} Operating on the physical layer
has benefits but also generates new challenges: we must aim to detect
any packet that might arrive at a network node, and ensure that all
violating packets are destroyed. These issues make the position and
number of guardians important factors during deployment of the guardian
system, and an optimization based on analytical models along the lines
of \prettyref{sec:Analysis} would be desirable. Methods that may
be applied for this purpose are presented in information theory literature
in the context of physical layer confidentiality in wireless networks
\cite{CarvingJam,ProtectJam}. These results suggest that by using
a security perimeter around the network attacks can be mitigated completely
because no network devices are then located in the attack range from
all reachable attacker locations. Alternatively, a training phase
or site survey can be helpful to support the optimization of the guardian
deployment.

\textbf{Alternative uses.} While the concept of protecting sensor
networks with the guardian system is our main focus, we note that
the generation of selective interference can also be useful for research
on the effects of interference on network performance, allowing to
perform repeatable experiments with real hardware. Related work in
this area uses sensor mote hardware for interference generation, which
limits the capabilities of such interference generation systems \cite{ReactiveTestbed,JamLab}.
Using the selective and protocol-aware interference generation capabilities
of the guardians, a more fine-grained control over interference can
be achieved. For example, the guardian system can be deployed in a
wireless testbed to generate arbitrary interference patterns based
on packet content. In this spirit, the guardians may enforce that
all ACK packets of Node~1 are lost, or 10\,\% of all network traffic
is affected by microwave\slash{}WLAN\slash{}Bluetooth-like interference
following a bursty pattern. We presented initial steps in this direction
recently \cite{SelectInterfere}.

\section{Related Work}

\label{sec:RelatedWork}
\begin{table*}
\centering\scriptsize %
\begin{tabular}{cccccc}
\toprule 
System & Application area & Maximum & Guard & Blocking criteria & Prototype\tabularnewline
\addlinespace[-1mm]
 &  & reaction time & distance &  & evaluation\tabularnewline
\midrule
\midrule 
IMD shield \cite{Heartbeats} & Implanted medical & \SI{10}{\milli\second} & 20\,cm & Each packet is blocked & $\surd$ (USRP2)\tabularnewline
 & devices (IMDs) &  &  & and selectively forwarded & \tabularnewline
\midrule
IMDGuard \cite{IMDGuard} & IMDs & Tens of ms & 20\,cm & Guard notices a spoofing attack & $\sim$ (MICAz)\tabularnewline
\midrule 
Warlock Duke \cite{LaShomb06} & Improvised explosives (IEDs) & n/a & n/a (100\,m) & Any signal in guarded freq.~bands & $\surd$ (custom)\tabularnewline
\midrule 
Blocker Tags \cite{BlockerTag,SoftBlocking} & RFID & \SI{300}{\micro\second} & 20\,cm & Tag query to protected prefix & $\times$ (tag)\tabularnewline
\midrule 
RFID Guardian \cite{RFIDGuardian,KeepBlocking} & RFID & \SI{300}{\micro\second} & 1\,m & Tag query to tag in ACL & $\times$ (handheld)\tabularnewline
\midrule 
Jamming for Good \cite{JamGood} & Sensor networks & \SI{5}{\milli\second} & 2--3\,m & Address+RSSI of registration packet & $\surd$ (MICAz)\tabularnewline
\midrule 
\textbf{{[}This work{]}, \cite{WMSL11-2}} & \textbf{Sensor networks} & \bfseries\SI{64}{\micro\second} & \textbf{10--20\,m} & \textbf{Per-packet decision (header+payload)} & \textbf{$\surd$ (USRP2)}\tabularnewline
\bottomrule
\end{tabular}

\caption{\label{tab:RelatedComparison}Comparison of related protection systems
using physical layer responses.}
\end{table*}
The concept of using selective interference has recently been proposed
in several application areas: to protect implanted medical devices
(IMDs) from malicious readers, to increase the privacy of RFID tags,
and to ensure authentic communication in WSNs. In contrast to these
works, we provide a system that allows configurable security policies
based on packet content and aims to provide a central protection over
larger distances in a \emph{networked} setting, in contrast to a reader--single
device setting. A summary of the following comparison is provided
in \prettyref{tab:RelatedComparison}.

IMDs face challenges similar to WSNs, namely low computational resources
and limited energy. Gollakota et al.~\cite{Heartbeats} describe
an external IMD protection system, or IMD ``shield,'' that allows
to regulate access to the device using selective interference, protecting
it from malicious readers. The shield is a battery-powered device
that is worn close the implanted device ($\sim20$\,cm), e.g., in
the form of a pendant. It acts as a proxy that simultaneously receives
and destroys any packet related to the protected IMD. If the packet
is going to the IMD, the shield checks whether the reader is trusted
and forwards the packet. If the packet originated from the IMD, the
packet is forwarded in encrypted form to the querying reader to protect
the patients's privacy. A USRP2-based prototype system is presented,
showing that an attacker can only succeed if it uses high transmit
power and close proximity. Xu et al.~\cite{IMDGuard} also describe
an external guardian system that protects IMDs from untrusted readers.
However, the concept relies on cryptographic protocols and uses the
physical layer response only as a countermeasure to spoofing attacks.
In contrast to these works, we offer configurable policies for several
devices in a distributed sensor network setting.

To protect the privacy of RFID tags from malicious readers, Juels
et al.~\cite{BlockerTag} introduce the ``blocker tag'' to prevent
the tag discovery by confusing readers with artificial collisions.
The attacker queries a prefix of node IDs (e.g., the first two bits),
and on collisions the reader refines the prefix, such that the blocker
tag can force the reader to traverse the full address space by generating
intentional interference. Juels and Brainard \cite{SoftBlocking}
extend this concept to signal privacy policies to benign readers.
Rieback et al.~\cite{RFIDGuardian,KeepBlocking} offer a similar
solution but support the protection of configurable sets of RFIDs
(blocker tags only support address blocks). A handheld device, the
``RFID Guardian,'' monitors all queries and interferes with a tag's
response to hide its presence from malicious readers. Again, these
concepts rely on close proximity to the protected tags. However, the
main difference to our work is that these schemes do not operate on
a per-packet basis: malicious queries are actually received by the
tag, and only the tag's response is blocked. This is not problematic
because RFID tags commonly do not keep state information. With our
implementation, we can prevent sensor motes from receiving any malicious
packet, also protecting their internal state.

A closely related work protects sensor motes from spoofed packets
(based on RSSI information) using selective interference \cite{JamGood,MGS08}.
Each data transmission is split in two packets, a registration packet
and a data packet. The protection is performed by the network motes
themselves, analyzing if the RSSI signature of the registration packet
matches with the claimed source address, and scheduling the transmission
of an interfering packet concurrent with the data packet in case of
a mismatch. However, the requirement that motes must receive packets
that are not addressed to them and send packets for interference is
expensive in terms of energy. So, while the goals are similar, the
approach is different. We explore the use of specially designed guardian
devices that provide a per-packet central enforcement of access policies,
without requiring a custom MAC protocol. 

In the context of firewalls and WSNs, there are also efforts to bring
an on-mote ``personal firewall'' to sensor networks \cite{AEGIS}.
Each incoming packet is inspected, compared to a set of filter rules,
and dropped from the receive queue if necessary. In contrast to this
work, we perform packet filtering using a centralized infrastructure,
lifting the requirement to distribute and manage the security configuration
of each node individually from a remote position.

In the military context, the U.S.~military employs mobile jamming
systems to protect convoys in Iraq from improvised roadside bombs,
stopping a bomb trigger signal from arriving at the bomb's receiver
\cite{LaShomb06}.%
\footnote{A news article on IED jammers used in the Iraq campaign is available
at \url{http://edition.cnn.com/2007/TECH/08/13/cied.jamming.tech}%
} However, not much is known about the system's implementation and
operation. 

In information theory the concept of \emph{secrecy capacity} of broadcast
channels with noise was studied extensively, starting with Wyner's
work on wiretap channels \cite{Wynertap}. The goal is to enable confidential
communications without secrets over a public broadcast channel. Recently,
several authors augmented the wiretap channel by considering the creation
of intentional noise to boost the secrecy capacity of the channel.
This approach is known under various designation, such as \emph{artificial
noise} \cite{ArtificialNoise,SecRateRegions}, \emph{cooperative jamming}
\cite{CoopJam,SecrecyTale}, \emph{friendly jamming} \cite{FriendlyJam},
the\emph{ relay-eavesdropper channel} \cite{RelaySecr}, or the \emph{wiretap
channel with helping interferer} \cite{InterferenceAssist}. These
works show that confidentiality in the sense of information theory
can be achieved even if the SINR of the adversary is higher. Our work
is orthogonal to these approaches because we exploit jamming to \emph{control}
medium access instead of ensuring message confidentiality at the physical
layer, i.e., while the jamming in the related work is targeting attackers
to prevent them to receive messages, our approach here is to jam the
nodes in the protected network, preventing them from receiving unauthorized
messages.

Work on wireless network security applies result on the wiretap channel
to prevent information leakage from a protected geographical area
by hindering eavesdroppers using intentional interference in more
practical scenarios. Kim et al.~\cite{CarvingJam} propose \emph{defensive
jamming,} a method to hinder eavesdroppers from detecting messages
correctly by using jamming directed towards the outside of the area.
The basic assumption of this approach is that the network deployment
area itself is physically secured, so that the eavesdropper must remain
outside a surrounding security perimeter. The authors perform an analysis
of jammer placement strategies based on the SINR model and show that
eavesdropping can be prevented successfully by an appropriate geometrical
jammer placement. The work of Sankararaman et al.~\cite{ProtectJam}
analyzes similar attack scenarios and presents algorithms for optimal
jamming power assignment and jammer placement, both for a fixed number
of jammers and a near-optimal number of jammers. While the methods
used on these works can be adapted to analyze the performance and
deployment of our guardian system, this work is orthogonal because
they consider passive (eavesdropping) attackers, while we focus on
active (injecting) attackers.

\section{Conclusion}

\label{sec:Conclusion}Wireless sensor networks are exposed to various
kinds of attacks that can be launched by exploiting the open access
characteristics of the physical broadcast medium. While cryptographic
techniques at higher layers may often counter these threats, real-world
networks often do not apply those techniques or have flawed implementations
that still open possible attack vectors to adversaries. We have presented
a guardian system operating on the physical layer as a practical solution
to solve the wireless channel access problem, and demonstrated the
feasibility of our approach with a full system implementation. Our
concept offers flexible solution with per-packet access rules and
the ability to enforce them from a central point in the network. Our
evaluation shows that our implementation fulfills the strict real-time
constraints when blocking high-throughput traffic with a reaction
time of \SI{39}{\micro\second} while achieving a blocking success
rate of over 99.9\,\%. We have further shown in several application
scenarios that the guardians prevent attacks that exploit the open
channel characteristics of wireless networks, and successfully reclaim
air dominance from the adversary.
\bibliographystyle{IEEEtran}
\bibliography{refs-long}

\end{document}